\documentclass[aps,prc,superscriptaddress,showpacs,nofootinbib,floatfix,onecolumn]{revtex4}
\usepackage{afterpage}
\usepackage{setspace}
\usepackage{rotating}
\usepackage{amssymb}
\usepackage{float}
\usepackage{harpoon}
\usepackage{amsmath}
\RequirePackage[colorlinks,pdfauthor={The ATLAS Collaboration}]{hyperref}  
\hypersetup{linkcolor=blue,citecolor=blue,filecolor=black,urlcolor=blue} 

\usepackage{atlasphysics} % Contains useful shortcuts. Uncomment to use
\usepackage{preprintcover}  % load package

\PreprintCoverPaperTitle{\boldmath Measurement of event-plane correlations in $\sqrt{s_{\mathrm{NN}}}=2.76$ TeV lead--lead collisions with the ATLAS detector}

\PreprintIdNumber{CERN-PH-EP-2014-021}  % CERN preprint number

\PreprintCoverAbstract{A measurement of event-plane correlations involving two or three event planes of different order is presented as a function of centrality for 7 $\mu$b$^{-1}$ Pb+Pb collision data at $\sqrt{s_{_{\mathrm{NN}}}}=2.76$ TeV, recorded by the ATLAS experiment at the LHC. Fourteen correlators are measured using a standard event-plane method and a scalar-product method, and the latter method is found to give a systematically larger correlation signal. Several different trends in the centrality dependence of these correlators are observed. These trends are not reproduced by predictions based on the Glauber model, which includes only the correlations from the collision geometry in the initial state. Calculations that include the final-state collective dynamics are able to describe qualitatively, and in some cases also quantitatively, the centrality dependence of the measured correlators. These observations suggest that both the fluctuations in the initial geometry and non-linear mixing between different harmonics in the final state are important for creating these correlations in momentum space.}  % paper abstract
\PreprintJournalName{Phys. Rev. C.}  % journal name
\begin{document}
% Title
\title{Measurement of event-plane correlations in $\sqrt{s_{\mathrm{NN}}}=2.76$ TeV lead--lead collisions with the ATLAS detector}
\author{The ATLAS Collaboration}
\begin{abstract}
A measurement of event-plane correlations involving two or three event planes of different order is presented as a function of centrality for 7 $\mu$b$^{-1}$ Pb+Pb collision data at $\sqrt{s_{_{\mathrm{NN}}}}=2.76$ TeV, recorded by the ATLAS experiment at the LHC. Fourteen correlators are measured using a standard event-plane method and a scalar-product method, and the latter method is found to give a systematically larger correlation signal. Several different trends in the centrality dependence of these correlators are observed. These trends are not reproduced by predictions based on the Glauber model, which includes only the correlations from the collision geometry in the initial state. Calculations that include the final-state collective dynamics are able to describe qualitatively, and in some cases also quantitatively, the centrality dependence of the measured correlators. These observations suggest that both the fluctuations in the initial geometry and non-linear mixing between different harmonics in the final state are important for creating these correlations in momentum space. 
\end{abstract}
\maketitle
\section{Introduction}
Heavy-ion collisions at the Relativistic Heavy Ion Collider (RHIC) and the Large Hadron Collider (LHC) create hot and dense matter that is thought to be composed of strongly interacting quarks and gluons. One striking observation that supports this picture is the large momentum anisotropy of particle emission in the transverse plane. This anisotropy is believed to be the result of anisotropic expansion of the created matter driven by the pressure gradients, with more particles emitted in the direction of the largest gradients~\cite{Voloshin:2008dg}. The collective expansion of the matter can be modeled by relativistic viscous hydrodynamic theory~\cite{Heinz:2013th}. The magnitude of the azimuthal anisotropy is sensitive to transport properties of the matter, such as the ratio of the shear viscosity to the entropy density and the equation of state~\cite{Teaney:2009qa}.

The anisotropy of the particle distribution ($\frac{dN}{d\phi}$) in azimuthal angle $\phi$ is customarily characterized by a Fourier series:
\begin{equation}
\frac{dN}{d\phi}\propto1+2\sum_{n=1}^{\infty}v_{n}\cos n(\phi-\Phi_{n})\;,
\end{equation}
where $v_n$ and $\Phi_n$ represent the magnitude and phase (referred to as the event plane) of the $n^{\mathrm{th}}$-order azimuthal anisotropy (or flow) at the corresponding angular scale. These quantities can also be conveniently represented in a two-dimensional vector format or in the standard complex form~\cite{Poskanzer:1998yz,Gardim:2011xv}:
\begin{equation} 
\label{eq:vec}
\overrightharp{v}_n=(v_n\cos n\Phi_{n},v_n\sin n\Phi_{n})\;\; {\mathrm {or}}\;\;v_n e^{i n\Phi_n}\;.
\end{equation}

In non-central collisions, the overlap region of the initial geometry has an almost elliptic shape. The anisotropy is therefore dominated by the second harmonic term, $v_2$. However, first-order ($n=1$) and higher-order ($n>2$) $v_n$ coefficients have also been observed~\cite{Aamodt:2011by,CMS:2012wg,Aad:2012bu}. These coefficients have been related to additional shape components arising from the fluctuations of the positions of nucleons in the overlap region. The amplitude and the directions of these shape components can be estimated via a simple Glauber model~\cite{Miller:2007ri} from the transverse positions $(r,\phi)$ of the participating nucleons relative to their center of mass~\cite{Teaney:2010vd}:
\begin{eqnarray}
\label{eq:ena}
\epsilon_n &=& \frac{\sqrt{\langle r^n\cos n\phi\rangle^2+\langle r^n\sin n\phi\rangle^2}}{\langle r^n\rangle},\\\label{eq:glau}
n\Phi_n^*&=&\arctan\left(\frac{\langle r^n\sin n\phi\rangle}{\langle r^n\cos n\phi\rangle}\right) +\pi\;,
\end{eqnarray} 
where $\epsilon_n$ is the eccentricity and the angle $\Phi^*_n$ is commonly referred to as the participant-plane (PP) angle. These shape components are transferred via hydrodynamic evolution into higher-order azimuthal anisotropy in momentum space. For small $\epsilon_n$ values, one expects $v_n\propto \epsilon_n$, and the $\Phi_n$ to be correlated with the minor-axis direction given by $\Phi_n^*$. However, model calculations show that the values of $\epsilon_n$ are large, and the alignment between $\Phi_n$ and $\Phi_n^*$ is strongly violated for $n>3$ due to non-linear effects in the hydrodynamic evolution~\cite{Qiu:2011iv}.

Detailed measurements of $v_n$ have been performed at RHIC and the LHC, and non-zero $v_n$ values are observed for $n\leq6$~\cite{Adare:2011tg,Adamczyk:2013waa,Aamodt:2010pa,Aamodt:2011by,Aad:2012bu,Aad:2013xma,CMS:2012wg,Chatrchyan:2013kba}, consistent with the existence of sizable fluctuations in the initial state. Further information on these fluctuations can be obtained by studying the correlations between $\Phi_{n}$ of different order. If the fluctuations in $\epsilon_n$ are small and totally random, the orientations of $\Phi_{n}$ of different order are expected to be uncorrelated. Calculations based on the Glauber model reveal strong correlations between some PP angles such as $\Phi_2^*$ and $\Phi_4^*$ or $\Phi_2^*$ and $\Phi_6^*$~\cite{Teaney:2010vd}, and weak correlations between others such as $\Phi_2^*$ and $\Phi_3^*$ or $\Phi_2^*$ and $\Phi_5^*$~\cite{Nagle:2010zk}. Previous measurements at RHIC and the LHC support a weak correlation between $\Phi_2$ and $\Phi_3$~\cite{ALICE:2011ab,Bhalerao:2011ry} and a strong correlation between $\Phi_2$ and $\Phi_4$~\cite{Adare:2010ux}. The former is consistent with no strong correlation between $\Phi_2^*$ and $\Phi_3^*$ and the dominance of linear response for elliptic flow and triangular flow, i.e. $\Phi_2\approx \Phi_2^*$ and $\Phi_3\approx \Phi_3^*$. The latter is consistent with a significant non-linear hydrodynamic response for quadrangular flow, which couples $v_4$ to $v_2^2$. The correlations among three event planes of different order have also been investigated in a model framework and several significant correlators have been identified~\cite{Teaney:2010vd,Bhalerao:2011yg,Jia:2012ma,Jia:2012ju}. However, no published experimental measurements on three-plane correlations exist to date. A measurement of the correlations between two and three event planes can shed light on the patterns of the fluctuations of the initial-state geometry and non-linear effects in the final state.

\section{ATLAS Detector and trigger}
The ATLAS detector~\cite{Aad:2008zzm} provides nearly full solid angle coverage of the collision point with tracking detectors, calorimeters and muon chambers, which are well suited for measurements of azimuthal anisotropies over a large pseudorapidity range.~\footnote{ATLAS uses a right-handed coordinate system with its origin at the nominal interaction point (IP) in the center of the detector and the $z$-axis along the beam pipe. The $x$-axis points from the IP to the center of the LHC ring, and the $y$-axis points upward. Cylindrical coordinates $(r,\phi)$ are used in the transverse plane, $\phi$ being the azimuthal angle around the beam pipe. The pseudorapidity is defined in terms of the polar angle $\theta$ as $\eta=-\ln\tan(\theta/2)$.} This analysis primarily uses three subsystems to measure the event plane: the inner detector (ID), the barrel and endcap electromagnetic calorimeters (ECal) and the forward calorimeter (FCal). The ID is contained within the 2~T field of a superconducting solenoid magnet, and measures the trajectories of charged particles in the pseudorapidity range $|\eta|<2.5$ and over the full azimuth. A charged particle passing through the ID typically traverses three modules of the silicon pixel detector (Pixel), four double-sided silicon strip modules of the semiconductor tracker (SCT), and a transition radiation tracker for $|\eta|<2$. The electromagnetic energy measurement in the ECal is based on a liquid-argon sampling technology. The FCal uses tungsten and copper absorbers with liquid argon as the active medium, and has a total thickness of about ten interaction lengths. The ECal covers the pseudorapidity range $|\eta|<3.2$, and the FCal extends the calorimeter coverage to $|\eta|< 4.9$. The energies in the ECal and FCal are reconstructed and grouped into towers with segmentation in pseudorapidity and azimuthal angle of $\Delta\eta\times\Delta\phi=0.1\times0.1$ to $0.2\times0.2$, which are then used to calculate the event plane. The procedure for obtaining the event-plane correlations is found to be insensitive to the segmentation and energy calibration of the calorimeters. 

The minimum-bias Level-1 trigger~\cite{Aad:2012xs} used for this analysis requires a signal in each of two zero-degree calorimeters (ZDC) or a signal in either one of the two minimum-bias trigger scintillator (MBTS) counters. The ZDC is positioned at 140~m from the collision point, detecting neutrons and photons with $|\eta|>8.3$, and the MBTS covers $2.1<|\eta|<3.9$. The ZDC Level-1 trigger thresholds on each side are set below the peak corresponding to a single neutron. A Level-2 timing requirement based on signals from each side of the MBTS is imposed to suppress beam backgrounds~\cite{Aad:2012xs}.

\section{Event and track selections}
\label{sec:sel}
This paper is based on Pb+Pb collision data collected in 2010 at the LHC with a nucleon--nucleon center-of-mass energy $\sqrt{s_{_{\mathrm{NN}}}}=2.76$~TeV. The data correspond to an integrated luminosity of approximately 7~$\mu\mathrm{b}^{-1}$. In order to suppress non-collision backgrounds, an offline event selection requires a reconstructed primary vertex with at least three associated charged tracks reconstructed in the ID and a time difference $|\Delta t| < 3$ ns between the MBTS trigger counters on either side of the interaction point. A coincidence between the two ZDCs at forward and backward pseudorapidity is required to reject a variety of background processes, while maintaining high efficiency for non-Coulomb processes. Events satisfying these conditions are required to have a reconstructed primary vertex with $z_{\mathrm{vtx}}$ within 150~mm of the nominal center of the ATLAS detector. The pile-up probability is estimated to be at the $10^{-4}$ level and is therefore negligible. About 48 million events pass the requirements for the analysis.

The Pb+Pb event centrality is characterized using the total transverse energy ($\Sigma \eT$) deposited in the FCal over the pseudorapidity range $3.2 < |\eta| < 4.9$ and measured at the electromagnetic energy scale~\cite{Aad:2011he}. A larger $\Sigma \eT$ value corresponds to a more central collision. From an analysis of the $\Sigma \eT$ distribution after applying all trigger and event selection criteria, the sampled fraction of the total inelastic cross section has been estimated to be (98$\pm$2)\% in a previous analysis~\cite{ATLAS:2011ag}.  The uncertainty in this estimate is evaluated by varying the trigger criteria, event selection and background rejection requirements on the FCal $\Sigma \eT$ distribution~\cite{ATLAS:2011ag}. The FCal $\Sigma \eT$ distribution is divided into a set of 5\%-wide percentile bins, together with five 1\%-wide bins for the most central 5\% of the events. A centrality interval refers to a percentile range, starting at 0\% for the most central collisions. Thus the 0\%--1\% centrality interval corresponds to the most central 1\% of the events. A standard Glauber model Monte Carlo analysis~\cite{Miller:2007ri} is used to estimate the average number of participating nucleons, $\langle N_{\mathrm{part}}\rangle$, and its associated systematic uncertainties for each centrality interval~\cite{ATLAS:2011ag}. These numbers are summarized in Table~\ref{tab:cent}.

\begin{table}[h!]
\centering
\begin{tabular}{|l|c|c|c|c|c|}\hline
Centrality               & 0\%--1\%        & 1\%--2\%        & 2\%--3\%       & 3\%--4\%      &4\%--5\%       \tabularnewline\hline
$\langle N_{\mathrm{part}}\rangle$ & $400.6\pm1.3$& $392.6\pm1.8$&$383.2\pm2.1$ &$372.6\pm2.3$ &$361.8\pm2.5$\tabularnewline\hline\hline
Centrality               & 0\%--5\%        & 5\%--10\%       & 10\%--15\%      &15\%--20\%   & 20\%--25\%       \tabularnewline\hline
$\langle N_{\mathrm{part}}\rangle$ & $382.2\pm2.0$&$330.3\pm3.0$ &$281.9\pm3.5$ &$239.5\pm3.8$& $202.6\pm3.9$\tabularnewline\hline\hline
Centrality               & 25\%--30\%      & 30\%--35\%      & 35\%--40\%      &40\%--45\% & 45\%--50\%           \tabularnewline\hline
$\langle N_{\mathrm{part}}\rangle$ & $170.2\pm4.0$&$141.7\pm3.9$ &$116.8\pm3.8$ &$95.0\pm3.7$& $76.1\pm3.5$\tabularnewline\hline\hline
Centrality               & 50\%--55\%      & 55\%--60\%      & 60\%--65\%      &65\%--70\%     & 70\%--75\% \tabularnewline\hline
$\langle N_{\mathrm{part}}\rangle$ & $59.9\pm3.3$  &$46.1\pm3.0$  &$34.7\pm2.7$  &$25.4\pm2.3$& $18\pm1.9$\tabularnewline\hline
\end{tabular}
\caption{\label{tab:cent} The list of centrality intervals and associated $\langle N_{\mathrm{part}}\rangle$ values used in this paper. The systematic uncertainties are taken from Ref.~\cite{ATLAS:2011ag}.}
\end{table}

The event plane is also measured by the ID, using reconstructed tracks with $\pT>0.5$ GeV and $|\eta|<2.5$~\cite{Aad:2012bu}. To improve the robustness of track reconstruction in the high-multiplicity environment of heavy-ion collisions, more stringent requirements on track quality, compared to those defined for proton--proton collisions~\cite{Aad:2010ac}, are used. At least nine hits in the silicon detectors are required for each track, with no missing Pixel hits and not more than one missing SCT hit, excluding the known non-operational modules. In addition, at its point of closest approach the track is required to be within 1~mm of the primary vertex in both the transverse and longitudinal directions~\cite{ATLAS:2011ah}. The track reconstruction performance is studied by comparing data to Monte Carlo calculations based on the HIJING event generator~\cite{Gyulassy:1994ew} and a full GEANT4 simulation of the detector~\cite{Aad:2010ah,Agostinelli:2002hh}. The track reconstruction efficiency ranges from 72\% at $\eta = 0$ to 51\% for $|\eta| > 2$ in peripheral collisions, while it ranges from 72\% at $\eta = 0$ to about 42\% for $|\eta| > 2$ in central collisions~\cite{trk}. However, the event-plane correlation results are found to be insensitive to the reconstruction efficiency (see Sec.~\ref{sec:sys}).

\section{Data analysis}
\label{sec:data}
\subsection{Experimental observables}
\label{sec:obs}

The $n^{\mathrm{th}}$-order harmonic has a $n$-fold symmetry in azimuth, and is thus invariant under the transformation $\Phi_n\rightarrow\Phi_n+2\pi/n$. Therefore, a general definition of the relative angle between two event planes, $a_n\Phi_n+a_m\Phi_m$, has to be invariant under a phase shift $\Phi_l\rightarrow\Phi_l+2\pi/l$. It should also be invariant under a global rotation by any angle. The first condition requires $a_n$ ($a_m$) to be multiple of $n$ ($m$), while the second condition requires the sum of the coefficients to vanish: $a_n+a_m=0$. The relative angle $\Phi_{n,m}=k(\Phi_n-\Phi_m)$, with $k$ being the least common multiple (LCM) of $n$ and $m$, satisfies these constraints, as does any integer multiple of $\Phi_{n,m}$. 

The correlation between $\Phi_n$ and $\Phi_m$ is completely described by the differential distribution of the event yield $dN_{\mathrm{evts}}/(d\left(k(\Phi_n-\Phi_m)\right)$. This distribution must be an even function due to the symmetry of the underlying physics, and hence can be expanded into the following Fourier series:
\begin{eqnarray}
\label{eq:o1c}
\frac{dN_{\mathrm{evts}}}{d\left(k(\Phi_n-\Phi_m)\right)}&\propto& 1+2\sum_{j=1}^{\infty} V_{n,m}^j \cos jk(\Phi_n-\Phi_m)\;,\\\label{eq:o1d}
V_{n,m}^j &=& \langle\cos jk(\Phi_n-\Phi_m)\rangle \;.
\end{eqnarray}
The measurement of the two-plane correlation is thus equivalent to measuring a set of cosine functions $\langle\cos jk(\Phi_n-\Phi_m)\rangle$ averaged over many events~\cite{Jia:2012ma}.

This discussion can be generalized for correlations involving three or more event planes. The multi-plane correlators can be written as $\langle\cos(c_1\Phi_{1}+2c_2\Phi_{2}...+lc_l\Phi_{l})\rangle$ with the constraint~\cite{Bhalerao:2011yg,Jia:2012ju}:
\begin{eqnarray}
\label{eq:o2a}
c_1+2c_2...+lc_l=0\;,
\end{eqnarray}
where the coefficients $c_n$ are integers. The two-plane correlators defined in Eq.~(\ref{eq:o1d}) satisfy this constraint. For convenience, correlation involving two event planes $\Phi_n$ and $\Phi_m$ is referred to as ``$n$-$m$'' correlation, and one involving three event planes $\Phi_n$, $\Phi_m$ and $\Phi_h$ as ``$n$-$m$-$h$'' correlation. The multi-plane correlators can always be decomposed into a linear combination of several two-plane relative angles and they carry additional information not accessible through two-plane correlators~\cite{Jia:2012ma}. 

Experimentally the $\Phi_n$ angles are estimated from the observed event-plane angles, $\Psi_{n}$, defined as the directions of the ``flow vectors'' $\overrightharp{q}_n$, which in turn are calculated from the azimuthal distribution of particles in the calorimeter or the ID:
\begin{eqnarray}
\nonumber
 \overrightharp{q}_n =(q_{x,n},q_{y,n}) &=& \frac{1}{\Sigma u_i}\left(\textstyle\Sigma [u_i\cos n\phi_i]-\langle\Sigma [u_i\cos n\phi_i]\rangle_{\mathrm{evts}}, \Sigma [u_i\sin n\phi_i]-\langle\Sigma [u_i\sin n\phi_i]\rangle_{\mathrm{evts}}\right)\;,\\\label{eq:2a} \tan n\Psi_n &=& \frac{q_{y,n}}{q_{x,n}}\;.
\end{eqnarray}
Here the weight $u_i$ is either the $\eT$ of the $i^{\mathrm{th}}$ tower in the ECal and the FCal or the $\pT$ of the $i^{\mathrm{th}}$ reconstructed track in the ID. Subtraction of the event-averaged centroid, $(\langle\Sigma [u_i\cos n\phi_i]\rangle_{\mathrm{evts}},\langle\Sigma [u_i\sin n\phi_i]\rangle_{\mathrm{evts}})$, in Eq.~(\ref{eq:2a}) removes biases due to detector effects~\cite{Afanasiev:2009wq}.~\footnote{For example, a localized inefficiency over a $\phi$ region in the detector would lead to a non-zero average $\protect\overrightharp{q}_n$. The subtraction corrects this bias.} A standard flattening technique~\cite{Barrette:1997pt} is then used to remove the small residual non-uniformities in the distribution of $\Psi_{n}$. The $\overrightharp{q}_n$ defined this way, when averaged over events with the same $\Phi_n$, is insensitive to the energy scale in the calorimeter or the momentum scale in the ID and to any random smearing effect. In the limit of infinite multiplicity it approaches the single-particle flow weighted by $u$: $\overrightharp{q}_n\rightarrow \left(\overrightharp{v}_n\right)_u= \Sigma u_i(\overrightharp{v}_n)_i/\Sigma u_i$.

The correlators in terms of $\Phi_n$ can be obtained from the correlations between the measured angles $\Psi_{n}$ divided by a resolution term~\cite{Jia:2012ma}:
\begin{eqnarray} 
\nonumber
\langle\cos (c_1\Phi_{1}+2c_2\Phi_{2}+...+lc_l\Phi_{l})\rangle &=& \frac{\langle\cos (c_1\Psi_{1}+2c_2\Psi_{2}+...+lc_l\Psi_{l})\rangle} {\mathrm{Res}\{c_1\Psi_1\}\mathrm{Res}\{c_22\Psi_2\}...\mathrm{Res}\{c_ll\Psi_l\}}\\\label{eq:o3a}
\mathrm{Res}\{c_nn\Psi_n\} &=& \sqrt{\left\langle\left(\cos c_nn(\Psi_n-\Phi_n)\right)^2\right\rangle}\;.
\end{eqnarray}
The resolution factors $\mathrm{Res}\{c_nn\Psi_n\}$ can be determined using the standard two-subevent or three-subevent methods~\cite{Poskanzer:1998yz} as discussed in Sec.~\ref{sec:ana}. To avoid auto-correlations, each $\Psi_n$ needs to be measured using subevents covering different $\eta$ ranges, preferably with a gap in between. Here a subevent refers to a collection of particles over a certain $\eta$ range in the event. This method of obtaining the correlator is referred to as the event-plane or EP method.

In Eq.~(\ref{eq:o3a}), all events are given equal weights in both the numerator (raw correlator) and the denominator (resolution). It was recently proposed~\cite{Luzum:2012da,Bhalerao:2013ina} that the potential bias in the EP method arising from the effects of event-by-event fluctuations of the flow and multiplicity can be removed by applying additional weight factors:
\begin{eqnarray} 
\nonumber
\langle\cos (c_1\Phi_{1}+2c_2\Phi_{2}+...+lc_l\Phi_{l})\rangle_{\mathrm{w}} &=& \frac{\langle\cos (c_1\Psi_{1}+2c_2\Psi_{2}+...+lc_l\Psi_{l})\rangle_{\mathrm{w}}} {\mathrm{Res}\{c_1\Psi_1\}_{\mathrm{w}}\mathrm{Res}\{c_22\Psi_2\}_{\mathrm{w}}...\mathrm{Res}\{c_ll\Psi_l\}_{\mathrm{w}}}\\\nonumber
\langle\cos (c_1\Psi_{1}+2c_2\Psi_{2}+...+lc_l\Psi_{l})\rangle_{\mathrm{w}} &=& \langle q_1^{c_1} q_2^{c_2}... q_l^{c_l}\cos (c_1\Psi_{1}+2c_2\Psi_{2}+...+lc_l\Psi_{l})\rangle \\\label{eq:o3ab}
\mathrm{Res}\{c_nn\Psi_n\}_{\mathrm{w}} &=& \sqrt{\left\langle \left({q_n^{c_n}\cos c_nn(\Psi_n-\Phi_n)}\right)^2\right\rangle}\;,
\end{eqnarray}
where the $q_n=|\overrightharp{q}_n|$ represents the magnitude of the flow vector of the subevent used to calculate the $\Psi_n$ (Eq.~(\ref{eq:2a})), and the subscript ``$\rm w$'' is used to indicate the $q_n$-weighting. This weighting method is often referred to as the ``scalar-product'' or SP method~\cite{Adler:2002pu}. Correspondingly, the weighted version of the PP correlators can be obtained by using the eccentricity $\epsilon_n$ defined in Eq.~(\ref{eq:ena}) as the weight~\cite{Bhalerao:2011yg}:
\begin{eqnarray} 
\label{eq:o3ac}
\langle\cos (c_1\Phi^*_{1}+2c_2\Phi^*_{2}+...+lc_l\Phi^*_{l})\rangle_{\mathrm{w}}= \frac{\langle \epsilon_1^{c_1} \epsilon_2^{c_2}... \epsilon_l^{c_l}\cos (c_1\Phi^*_{1}+2c_2\Phi^*_{2}+...+lc_l\Phi^*_{l})\rangle} {\sqrt{\langle\epsilon_1^{2c_1}\rangle\langle\epsilon_2^{2c_2}\rangle...\langle\epsilon_l^{2c_l}\rangle}}\;.
\end{eqnarray}

In Eq.~(\ref{eq:o3ab}), events with larger flow have bigger weights in the calculation of the raw correlation and the resolution factors. Other than the weighting, the procedure for obtaining the raw signal and resolution factors is identical in the EP and SP methods. Hence the discussion in the remainder of the paper should be regarded as applicable to both methods and the subscript ``$\rm w$'' is dropped in all formulae, unless required for clarity.

It is worth emphasizing that the expression for the correlators in Eq.~\ref{eq:o3ab} is constructed to be insensitive to the details of the detector performance, such as the $\eta$-coverage, segmentation, energy calibration or the efficiency~\cite{Poskanzer:1998yz, Afanasiev:2009wq}. This is because the angle $\Phi_n$ is a global property of the event that can be estimated from the $\Psi_n$ from independent detectors, and the procedure for obtaining the correlators is ``self-correcting''. A poor segmentation or energy calibration of the calorimeter, for example, increases the smearing of $\Psi_n$ about $\Phi_n$, and hence reduces the raw correlation (numerator of Eq.~\ref{eq:o3ab}). This reduction in the raw correlation, however, is expected to be mostly compensated by smaller resolution terms $\mathrm{Res}\{c_nn\Psi_n\}$ in the denominators.

A very large number of correlators could be studied. However, the measurability of these correlators is dictated by the values of ${\mathrm{Res}\{jn\Psi_{n}\}}$ ($c_n$ replaced by $j$ for simplicity). A detailed study in this analysis shows that the values of ${\mathrm{Res}\{jn\Psi_{n}\}}$ decrease very quickly for increasing $n$, but they decrease more slowly with $j$ for fixed $n$~\cite{Jia:2012ju}. The resolution factors are sufficiently good for ${\mathrm{Res}\{jn\Psi_{n}\}}$ for $n=2$ to 6 and $j$ values up to $j=6$ for $n=2$. This defines the two- and three-plane correlators that can be measured.    
%These dependences limit the types of two- and three-plane correlations that can be measured in practice. This analysis calculates ${\mathrm{Res}\{jn\Psi_{n}\}}$ for $n=2$ to 6 and $j$ values up to $j=6$ for $n=2$.

Table~\ref{tab:2} gives a summary of the set of two-plane correlators and resolution terms that need to be measured in this analysis for each centrality interval. The corresponding information for the three-plane correlators is shown in Table~\ref{tab:3}. The first three correlators in Table~\ref{tab:2} correspond to the first three Fourier coefficients ($j=$1,2,3) in Eq.~(\ref{eq:o1d}), and are derived from the observed distribution $dN_{\mathrm{evts}}/d\left(4(\Psi_2-\Psi_4)\right)$. All the other correlators in Tables~\ref{tab:2} and ~\ref{tab:3} only correspond to the first Fourier coefficient of the observed distribution. The two-plane and three-plane correlators are listed separately because different subdetectors are used (see Sec.~\ref{sec:ana}), and this requires separate evaluation of the resolution corrections.
\begin{table}[h]
\centering
\begin{tabular}{|c|c|}\hline
$\langle\cos 4(\Phi_2-\Phi_4)\rangle$ &  $\mathrm{Res}\{4\Psi_{2}\}$,  $\mathrm{Res}\{4\Psi_{4}\}$\\
$\langle\cos 8(\Phi_2-\Phi_4)\rangle$ &  $\mathrm{Res}\{8\Psi_{2}\}$,  $\mathrm{Res}\{8\Psi_{4}\}$\\
$\langle\cos 12(\Phi_2-\Phi_4)\rangle$&  $\mathrm{Res}\{12\Psi_{2}\}$, $\mathrm{Res}\{12\Psi_{4}\}$\\
$\langle\cos 6(\Phi_2-\Phi_3)\rangle$ &  $\mathrm{Res}\{6\Psi_{2}\}$,  $\mathrm{Res}\{6\Psi_{3}\}$\\
$\langle\cos 6(\Phi_2-\Phi_6)\rangle$ &  $\mathrm{Res}\{6\Psi_{2}\}$,  $\mathrm{Res}\{6\Psi_{6}\}$\\
$\langle\cos 6(\Phi_3-\Phi_6)\rangle$ &  $\mathrm{Res}\{6\Psi_{3}\}$,  $\mathrm{Res}\{6\Psi_{6}\}$\\
$\langle\cos 12(\Phi_3-\Phi_4)\rangle$ &  $\mathrm{Res}\{12\Psi_{3}\}$,  $\mathrm{Res}\{12\Psi_{4}\}$\\
$\langle\cos 10(\Phi_2-\Phi_5)\rangle$ &  $\mathrm{Res}\{10\Psi_{2}\}$,  $\mathrm{Res}\{10\Psi_{5}\}$\\\hline
\end{tabular}
\caption{\label{tab:2} The list of two-plane correlators and associated event-plane resolution factors that need to be measured.}
\end{table}

\begin{table}[h]
\centering
\begin{tabular}{|c|c|}\hline
$\langle\cos (2\Phi_2+3\Phi_3-5\Phi_5)\rangle$ &  $\mathrm{Res}\{2\Psi_{2}\}$,  $\mathrm{Res}\{3\Psi_{3}\}$,  $\mathrm{Res}\{5\Psi_{5}\} $\\
$\langle\cos (-8\Phi_2+3\Phi_3+5\Phi_5)\rangle$ &  $\mathrm{Res}\{8\Psi_{2}\}$,  $\mathrm{Res}\{3\Psi_{3}\}$,  $\mathrm{Res}\{5\Psi_{5}\} $\\
$\langle\cos (2\Phi_2+4\Phi_4-6\Phi_6)\rangle$ &  $\mathrm{Res}\{2\Psi_{2}\}$,  $\mathrm{Res}\{4\Psi_{4}\}$,  $\mathrm{Res}\{6\Psi_{6}\} $\\
$\langle\cos (-10\Phi_2+4\Phi_4+6\Phi_6)\rangle$ &  $\mathrm{Res}\{10\Psi_{2}\}$,  $\mathrm{Res}\{4\Psi_{4}\}$,  $\mathrm{Res}\{6\Psi_{6}\} $\\
$\langle\cos (2\Phi_2-6\Phi_3+4\Phi_4)\rangle$ &  $\mathrm{Res}\{2\Psi_{2}\}$,  $\mathrm{Res}\{6\Psi_{3}\}$,  $\mathrm{Res}\{4\Psi_{4}\} $\\
$\langle\cos (-10\Phi_2+6\Phi_3+4\Phi_4)\rangle$ &  $\mathrm{Res}\{10\Psi_{2}\}$,  $\mathrm{Res}\{6\Psi_{3}\}$,  $\mathrm{Res}\{4\Psi_{4}\} $\\\hline
\end{tabular}
\caption{\label{tab:3} The list of three-plane correlators and associated event-plane resolution factors that need to be measured.}
\end{table}

\subsection{Analysis method}
\label{sec:ana}
For two-plane correlation (2PC) measurements, the event is divided into two subevents symmetric around $\eta=0$ with a gap in between, so they nominally have the same resolution. Each subevent provides its own estimate of the event plane via Eq.~(\ref{eq:2a}): $\Psi_n^{\mathrm{P}}$ and $\Psi_m^{\mathrm{P}}$ for positive $\eta$ and $\Psi_n^{\mathrm{N}}$ and $\Psi_m^{\mathrm{N}}$ for negative $\eta$. This leads to two statistically independent estimates of the correlator, which are averaged to obtain the final signal. Because of the symmetry of the subevents, the product of resolution factors in the denominator is identical for each measurement, and the event-averaged correlator can be written as:
\begin{eqnarray}
\label{eq:2db}
 \langle\cos k(\Phi_n-\Phi_m)\rangle= \frac{\left\langle\cos k(\Psi_n^{\mathrm{P}}-\Psi_m^{\mathrm{N}})\right\rangle+\left\langle\cos k(\Psi_n^{\mathrm{N}}-\Psi_m^{\mathrm{P}})\right\rangle}{\mathrm{Res}\{k\Psi_n^{\mathrm{P}}\}\mathrm{Res}\{k\Psi_m^{\mathrm{N}}\}+\mathrm{Res}\{k\Psi_n^{\mathrm{N}}\}\mathrm{Res}\{k\Psi_m^{\mathrm{P}}\}}\;.
\end{eqnarray}

To measure a three-plane correlation (3PC), three non-overlapping subevents, labeled as A, B and C, are chosen to have approximately the same $\eta$ coverage. In this analysis, subevents A and C are chosen to be symmetric about $\eta=0$, and hence have identical resolution, while the resolution of subevent B in general is different. There are $3!=6$ independent ways of obtaining the same three-plane correlator. But the symmetry between A and C reduces this to three pairs of measurements, which are labeled as Type1, Type2 and Type3. For example, the Type1 measurement of the correlation $2\Phi_2+3\Phi_3-5\Phi_5$ is obtained from $2\Psi_2^{\mathrm{B}}+3\Psi_3^{\mathrm{A}}-5\Psi_5^{\mathrm{C}}$ and $2\Psi_2^{\mathrm{B}}+3\Psi_3^{\mathrm{C}}-5\Psi_5^{\mathrm{A}}$, i.e. by requiring the $\Psi_2$ angle to be given by subevent B:
\begin{eqnarray}
\label{eq:2e}
\left\langle\cos (2\Phi_2+3\Phi_3-5\Phi_5)\right\rangle_{\mathrm{Type1}}=\frac{\left\langle\cos (2\Psi_2^{\mathrm{B}}+3\Psi_3^{\mathrm{A}}-5\Psi_5^{\mathrm{C}})\right\rangle+\left\langle\cos  (2\Psi_2^{\mathrm{B}}+3\Psi_3^{\mathrm{C}}-5\Psi_5^{\mathrm{A}})\right\rangle} {\mathrm{Res}\{2\Psi_2^{B}\}\mathrm{Res}\{3\Psi_3^{A}\}\mathrm{Res}\{5\Psi_5^{C}\}+\mathrm{Res}\{2\Psi_2^{B}\}\mathrm{Res}\{3\Psi_3^{C}\}\mathrm{Res}\{5\Psi_5^{A}\}}\;.
\end{eqnarray}
Similarly, the Type2 (Type3) measurement is obtained by requiring the $\Psi_3$ ($\Psi_5$) to be measured by subevent B. Since the three angles in each detector, e.g. $\Psi^{\mathrm{A}}_2$, $\Psi^{\mathrm{A}}_3$ and $\Psi^{\mathrm{A}}_5$, are obtained from orthogonal Fourier modes, the different types of estimates for a given correlator are expected to be statistically independent. 

The resolution factors $\mathrm{Res}\{jn\Psi_{n}\}$ are obtained from a two-subevent method (2SE) and a three-subevent method (3SE)~\cite{Poskanzer:1998yz}. The 2SE method follows almost identically the 2PC procedure described above: two subevents symmetric about $\eta=0$ are chosen and used to make two measurements of the event plane at the same order $n$: $\Psi_n^{\mathrm{P}}$ and $\Psi_n^{\mathrm{N}}$. The correlator $\langle \cos jn(\Psi_n^{\mathrm P}-\Psi_n^{\mathrm N})\rangle$ is then calculated, and the square-root yields the desired resolution~\cite{Aad:2012bu}:
\begin{eqnarray}
\label{eq:mep3}
\mathrm{Res}\{jn\Psi_{n}\} = \sqrt{\langle \cos jn(\Psi_n^{\mathrm P}-\Psi_n^{\mathrm N})\rangle}\equiv\mathrm{Res}\{jn\Psi_n^{\mathrm{P}}\} \equiv \mathrm{Res}\{jn\Psi_n^{\mathrm{N}}\}\;.
\end{eqnarray}

In the 3SE method, the value of $\mathrm{Res}\{jn\Psi_{n}\}$ for a given subevent A is determined from angle correlations with two subevents B and C covering different regions in $\eta$:
\begin{eqnarray}
\nonumber
{\mathrm{Res}}\{jn\Psi^{\mathrm A}_{n}\}=\sqrt{\frac{\left\langle {\cos jn \left(\Psi_n^{\mathrm A} - \Psi_n^{\mathrm B}\right)} \right\rangle\left\langle {\cos jn \left(\Psi_n^{\mathrm A} - \Psi_n^{\mathrm C}\right)} \right\rangle}{\left\langle {\cos jn \left(\Psi_n^{\mathrm B} - \Psi_n^{\mathrm C}\right)} \right\rangle}}.\\\label{eq:mep4}
\end{eqnarray}
The 3SE method does not rely on equal resolutions for the subevents, and hence there are many ways of choosing subevents B and C.

In the case of the weighted correlators given by the SP method, the resolution terms defined by Eqs.~\ref{eq:mep3} and \ref{eq:mep4} are instead calculated as~\cite{Luzum:2012da}:
\begin{eqnarray}
\label{eq:mep3b}
\mathrm{Res}\{jn\Psi_{n}\}_{\mathrm{w}}=\sqrt{\langle (q_n^{\mathrm P}q_n^{\mathrm N})^j\cos jn(\Psi_n^{\mathrm P}-\Psi_n^{\mathrm N})\rangle}\;,
\end{eqnarray}
and
\begin{eqnarray}
\label{eq:mep4b}
{\mathrm{Res}}\{jn\Psi^{\mathrm A}_{n}\}_{\mathrm{w}}= \sqrt{\frac{\left\langle { (q_n^{\mathrm A}q_n^{\mathrm B})^j \cos jn \left(\Psi_n^{\mathrm A} - \Psi_n^{\mathrm B}\right)} \right\rangle\left\langle { (q_n^{\mathrm A}q_n^{\mathrm C})^j\cos jn \left(\Psi_n^{\mathrm A} - \Psi_n^{\mathrm C}\right)} \right\rangle}{\left\langle { (q_n^{\mathrm B}q_n^{\mathrm C})^j\cos jn \left(\Psi_n^{\mathrm B} - \Psi_n^{\mathrm C}\right)} \right\rangle}}\;.
\end{eqnarray}

\subsection{Analysis procedure}
\label{sec:step}

The large $\eta$ coverage of the ID, ECal and FCal, with their fine segmentation, allows many choices of subevents for estimating the event planes and studying their correlations over about ten units in $\eta$. The edge towers of the FCal (approximately $4.8<|\eta|<4.9$) are excluded to minimize the non-uniformity of $\eT$ in azimuth, as in a previous analysis~\cite{Aad:2012bu}. These detectors are divided into a set of small segments in $\eta$, and the subevents are constructed by combining these segments. A large number of subevents can be used for measuring both the raw correlation signal and the resolution corrections. A detailed set of cross-checks and estimations of systematic uncertainties can therefore be performed. 

The guiding principle for choosing the subevents is that they should have large $\eta$ acceptance, but still have a sufficiently large $\eta$ gap from each other. For two-plane correlations, the default subevents are ECal+FCal at negative ($-4.8<\eta<-0.5$) and positive ($0.5<\eta<4.8$) $\eta$, with a gap of one unit in between. For three-plane correlations, the default subevents are ECal$_{\mathrm{P}}$ ($0.5<\eta<2.7$), FCal ($3.3<|\eta|<4.8$), and ECal$_{\mathrm{N}}$ ($-2.7<\eta<-0.5$). As an important consistency cross-check for the 2PC and 3PC analyses, subevents are also chosen only from the ID. These combinations are listed in Table~\ref{tab:dets}. The resolution for each of these subevents is determined via the 2SE method and the 3SE method, and the latter typically involves measuring correlations with many other subevents not listed in Table~\ref{tab:dets}, for example using smaller sections of the ECal or ID.

\begin{table}[h]
\centering
\begin{tabular}{|c||r|r|r|r|r|r|}\hline 
\multicolumn{7}{|c|}{Subevents used for two-plane correlations and their $\eta$ coverages}\tabularnewline\hline
Calorimeter-based     &\multicolumn{3}{|r|}{\hspace{0.8cm}ECalFCal$_{\mathrm{P}}$ $\eta\in$(0.5,4.8)} & \multicolumn{3}{|r|}{ECalFCal$_{\mathrm{N}}$ $\eta\in$($-$4.8,$-$0.5)}\tabularnewline\hline 
ID-based &\multicolumn{3}{|r|}{ID$_{\mathrm{P}}$ $\eta\in$(0.5,2.5)}       & \multicolumn{3}{|r|}{ID$_{\mathrm{N}}$ $\eta\in$($-$2.5,$-$0.5)}       \tabularnewline\hline
\multicolumn{7}{|c|}{Subevents used for three-plane correlations and their $\eta$ coverages}\tabularnewline\hline
Calorimeter-based    &\multicolumn{2}{|r|}{ECal$_{\mathrm{P}}$ $\eta\in$(0.5,2.7)}  & \multicolumn{2}{|r|}{FCal $|\eta|\in$(3.3,4.8)}   & \multicolumn{2}{|r|}{ECal$_{\mathrm{N}}$ $\eta\in$($-$2.7,$-$0.5)}\tabularnewline\hline
ID-based &\multicolumn{2}{|r|}{ID$_{\mathrm{P}}$ $\eta\in$(1.5,2.5) }   & \multicolumn{2}{|r|}{ID  $\eta\in$($-$1.0,1.0)}         & \multicolumn{2}{|r|}{ID$_{\mathrm{N}}$ $\eta\in$($-$2.5,$-$1.5)}  \tabularnewline\hline
\end{tabular}\caption{\label{tab:dets} Combinations of subevents used in two-plane and three-plane correlation analysis. The calorimter-based analysis is the default, while the ID-based result provides an important cross-check.}
\end{table}

\begin{figure}[h!]
\begin{center}
\includegraphics[width=1\columnwidth]{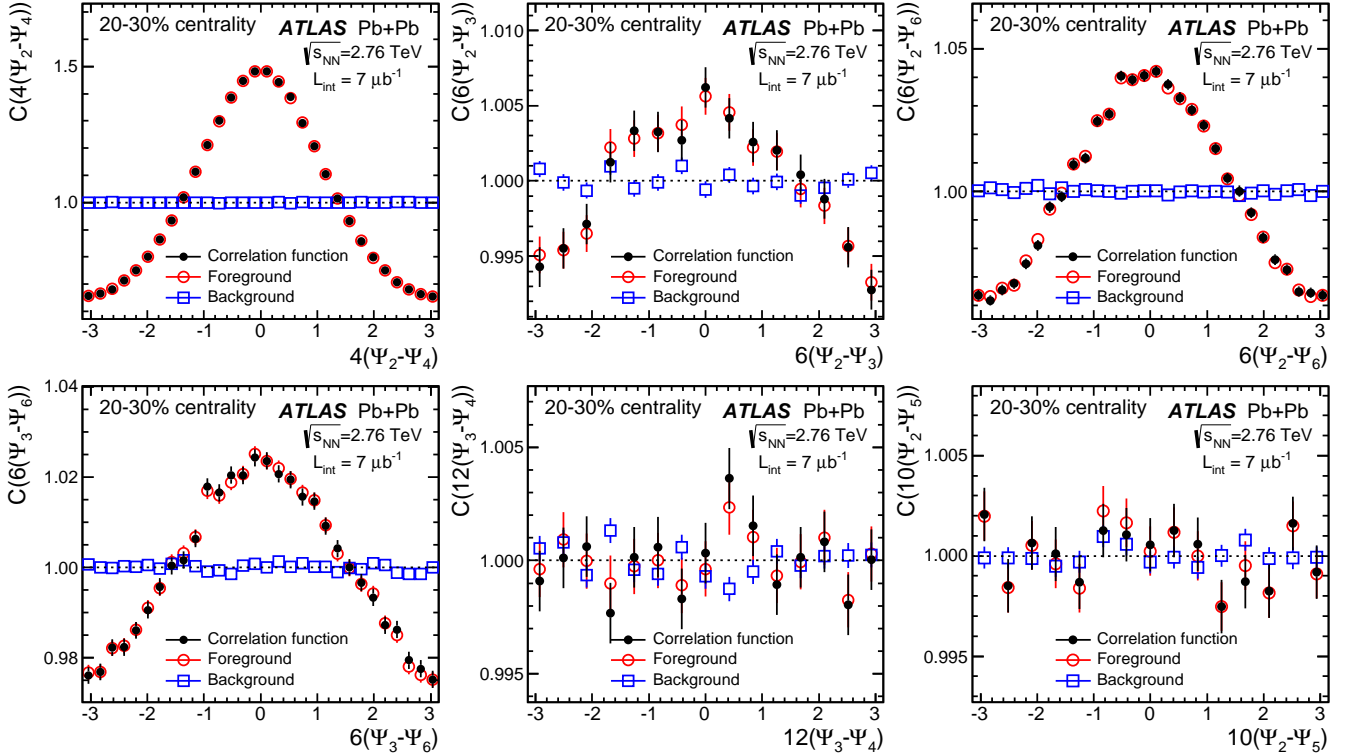}
\end{center}
\caption{\label{fig:2pcraw} (Color online) Relative angle distributions between two raw event planes from ECalFCal$_{\mathrm{N}}$ and ECalFCal$_{\mathrm{P}}$ defined in Table~\ref{tab:dets} for the 20\%--30\% centrality interval for the foreground (open circles), background (open squares) and correlation function (filled circles) based on the EP method. The correlation functions give (via Eq.~(\ref{eq:2db})) the two-plane correlators defined in Table~\ref{tab:2}. The $y$-axis scales are not the same for all panels.}
\end{figure}

Figure~\ref{fig:2pcraw} shows the two-plane relative angle distributions for the 20\%--30\% centrality interval. The signal or ``foreground'' distributions are calculated by combining event-plane angles from the same event. The ``background'' distributions are calculated from mixed events by combining the event-plane angles obtained from different events with similar centrality (matched within 5\%) and $z_{\mathrm{vtx}}$ (matched within 3~cm). Ten mixed events are constructed for each foreground event. Both distributions are normalized so that the average of the entries is one. The background distributions provide an estimate of detector effects, while the foreground distributions contain both the detector effects and physics. The background distributions are almost flat, but do indicate some small variations at a level of about $10^{-3}$. To cancel these non-physical structures, the correlation functions are obtained by dividing the foreground (S) by the background distributions (B):
\begin{eqnarray}
\label{eq:cf1}
C(k(\Psi_{n}-\Psi_{m})) &=& \frac{S(k(\Psi_{n}-\Psi_{m}))}{B(k(\Psi_{n}-\Psi_{m}))}\;.
\end{eqnarray}
The correlation functions show significant positive signals for $4(\Psi_2-\Psi_4)$, $6(\Psi_2-\Psi_3)$, $6(\Psi_2-\Psi_6)$ and $6(\Psi_3-\Psi_6)$. The observed correlation signals (not corrected by resolution) in terms of the cosine average are calculated directly from these correlation functions.

\begin{figure}[h!]
\begin{center}
\includegraphics[width=1\columnwidth]{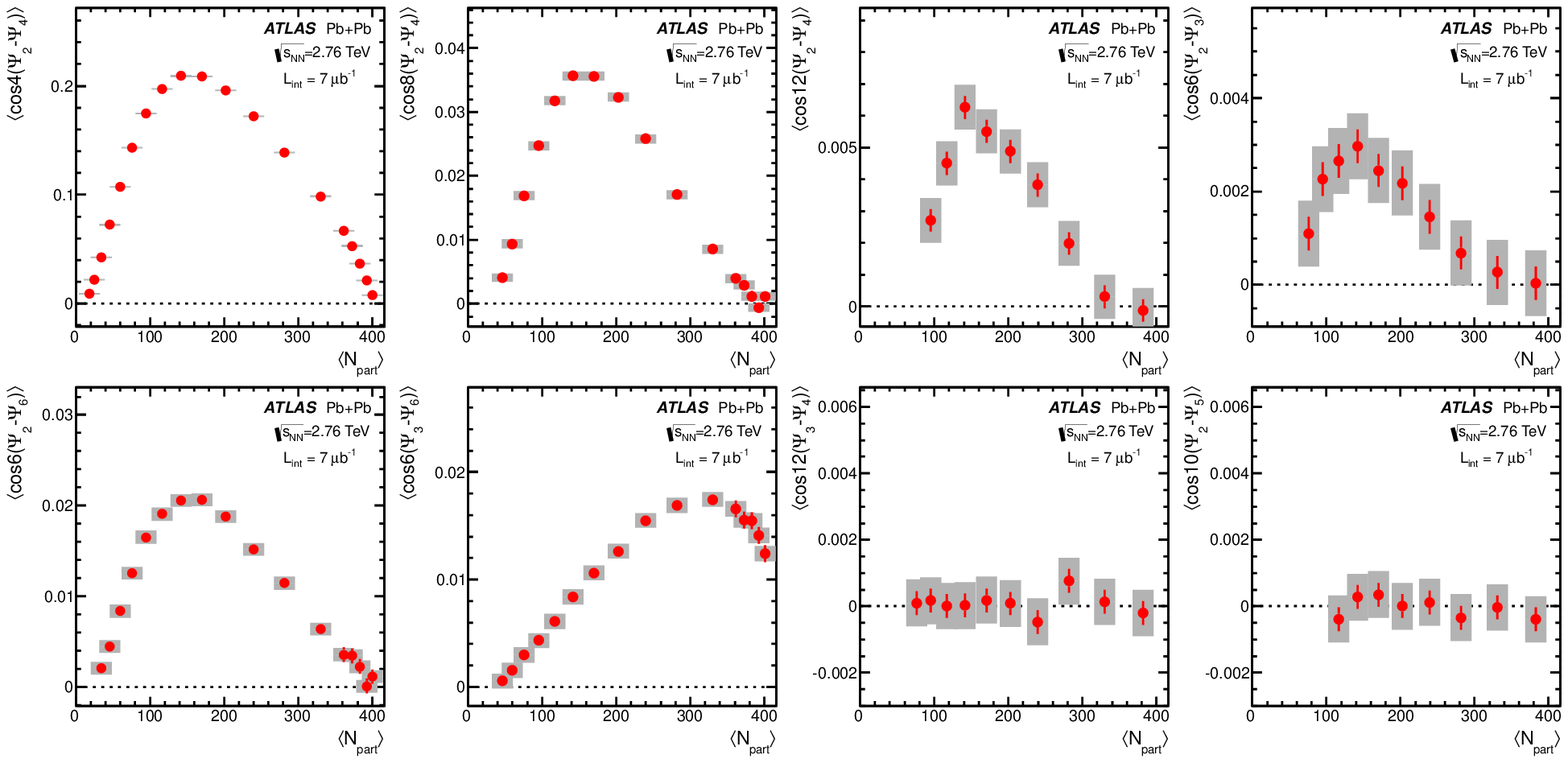}
\end{center}
\caption{\label{fig:raw1} (Color online) Observed correlation signals based on the EP method, $\langle\cos jk(\Psi_n-\Psi_m)\rangle$, calculated from the correlation functions such as those in Fig.~\ref{fig:2pcraw} as a function of $\langle N_{\mathrm{part}}\rangle$. The middle two panels in the top row have $j=2$ and $j=3$, while all other panels have $j=1$. The error bars and shaded bands indicate the statistical and systematic uncertainties, respectively.}
\end{figure}

Figure~\ref{fig:raw1} shows the centrality dependence of the observed correlation signals for various two-plane correlators. The systematic uncertainty, shown as shaded bands, is estimated as the values of the sine terms $\left\langle \sin jk\left(\Psi_n-\Psi_m\right)\right\rangle$. Non-zero sine terms may arise from detector effects which lead to non-physical correlations between the two subevents. This uncertainty is calculated by averaging sine terms across the measured centrality range, giving uncertainties of (0.2--1.5)$\times10^{-3}$ depending on the type of the correlator. This uncertainty is correlated with centrality and is significant only when the $\left\langle \cos jk\left(\Psi_n-\Psi_m\right)\right\rangle$ term is itself small, as in the rightmost four panels of Fig.~\ref{fig:raw1}. This uncertainty is included in the final results (see Sec.~\ref{sec:sys}).

A large number of resolution factors $\mathrm{Res}\{jn\Psi_{n}\}$ needs to be determined using the 2SE and the 3SE methods, separately for each subevent listed in Table~\ref{tab:dets}. For example, the resolution of ECalFCal$_{\mathrm{N}}$ can be obtained from its correlation with ECalFCal$_{\mathrm{P}}$ via Eq.~(\ref{eq:mep3}) (2SE method), or from its correlation with any two non-overlapping reference subevents at $\eta>0$ via Eq.~(\ref{eq:mep4}) (3SE method) such as $0.5<\eta<1.5$ and $3.3<\eta<4.8$. Therefore, for a particular subevent in Table~\ref{tab:dets}, there are usually several determinations of $\mathrm{Res}\{jn\Psi_{n}\}$, one from the 2SE method and several from the 3SE method. The default value used is obtained from the 2SE method where available, or from the 3SE combination with the smallest uncertainty. The spread of these values is included in the systematic uncertainty, separately for each centrality interval. The relative differences between most of these estimates are found to be independent of the event centrality, except for the 50\%--75\% centrality range, where weak centrality dependences are observed in some cases.

\begin{figure}[t]
\begin{center}
\includegraphics[width=1\columnwidth]{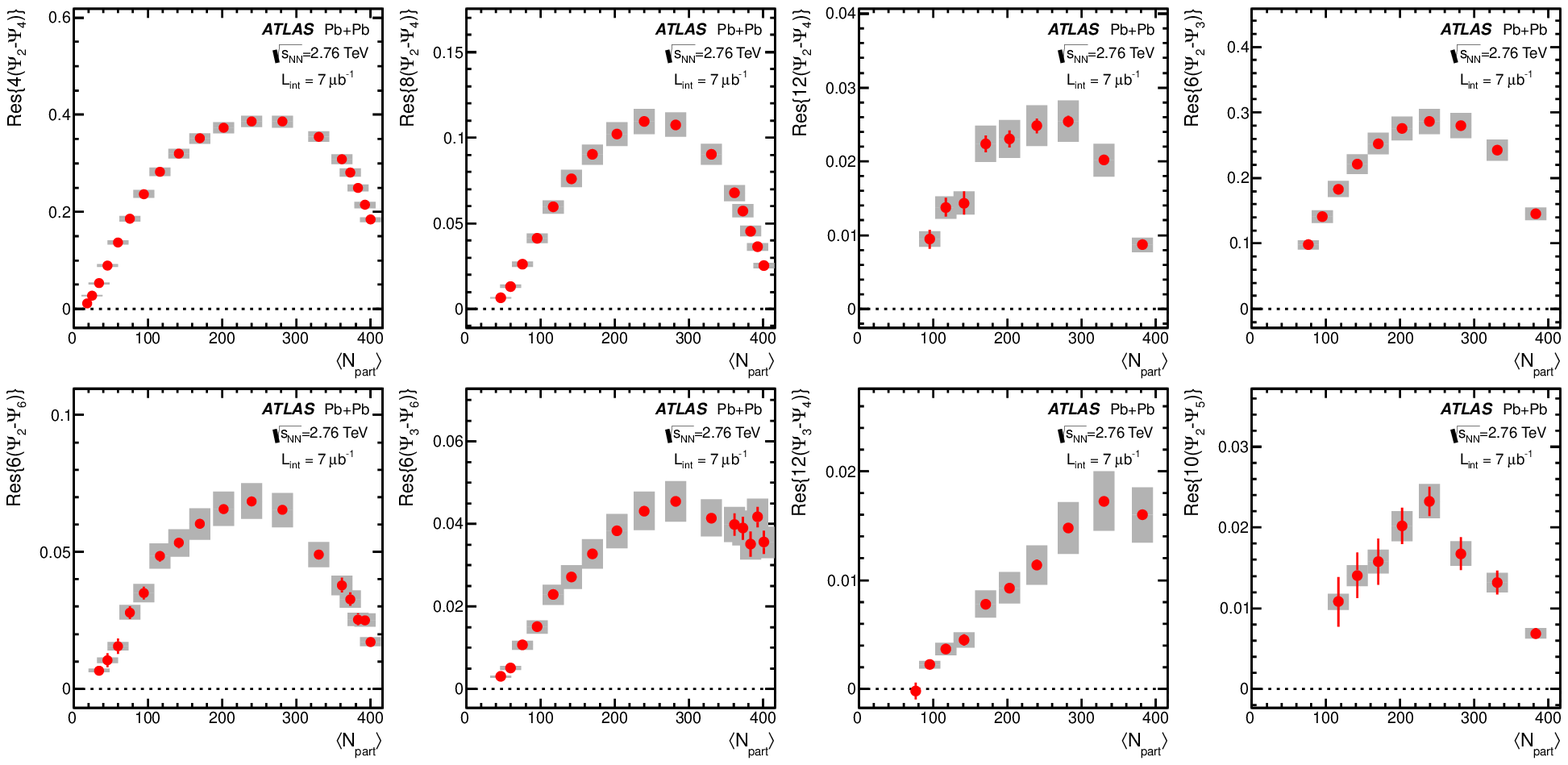}
\end{center}
\caption{\label{fig:reso1} (Color online) The combined resolution factors based on the EP method for two-plane correlators, Res$\{jk(\Psi_n-\Psi_{m})\}\equiv\mathrm{Res}\{jk\Psi_n\}\mathrm{Res}\{jk\Psi_m\}$, as a function of $\langle N_{\mathrm{part}}\rangle$. The middle two panels in the top row have $j=2$ and $j=3$, while all other panels have $j=1$. The error bars and shaded bands indicate the statistical and systematic uncertainties, respectively.}
\end{figure}

All the cosine terms in the 2SE and 3SE formulae are calculated from the distributions similar to those in Eq.~(\ref{eq:cf1}), but at the same order $n$:
\begin{eqnarray}
  \label{eq:cf1b}
  C(n(\Psi_{n}^{\mathrm{A}}-\Psi_{n}^{\mathrm{B}}))=\frac{S(n(\Psi_{n}^{\mathrm{A}}-\Psi_{n}^{\mathrm{B}}))}{B(n(\Psi_{n}^{\mathrm{A}}-\Psi_{n}^{\mathrm{B}}))}, 
\end{eqnarray}
where the background distribution is obtained by combining the $\Psi_{n}$ of subevent A in one event with $\Psi_{n}$ of subevent B from a different event with similar centrality and $z_{\mathrm{vtx}}$. Furthermore, the non-zero sine values $\left\langle \sin jn\left(\Psi_n^{\mathrm{A}}-\Psi_n^{\mathrm{B}}\right)\right\rangle$ arising from the 2SE and 3SE analyses are also included in the uncertainty in the resolution factor. Once the individual resolution factors are determined for each subevent, the combined resolution factors are then calculated by multiplying the relevant individual $\mathrm{Res}\{jn\Psi_{n}\}$ terms. They are shown in Fig.~\ref{fig:reso1} as a function of centrality for the eight two-plane correlators listed in Table~\ref{tab:2}. The systematic uncertainty is calculated via a simple error propagation from the individual resolution terms, and is nearly independent of the event centrality. This uncertainty is included in the final results (see Sec.~\ref{sec:sys}).

\begin{figure}[h!]
\begin{center}
\includegraphics[width=1\columnwidth]{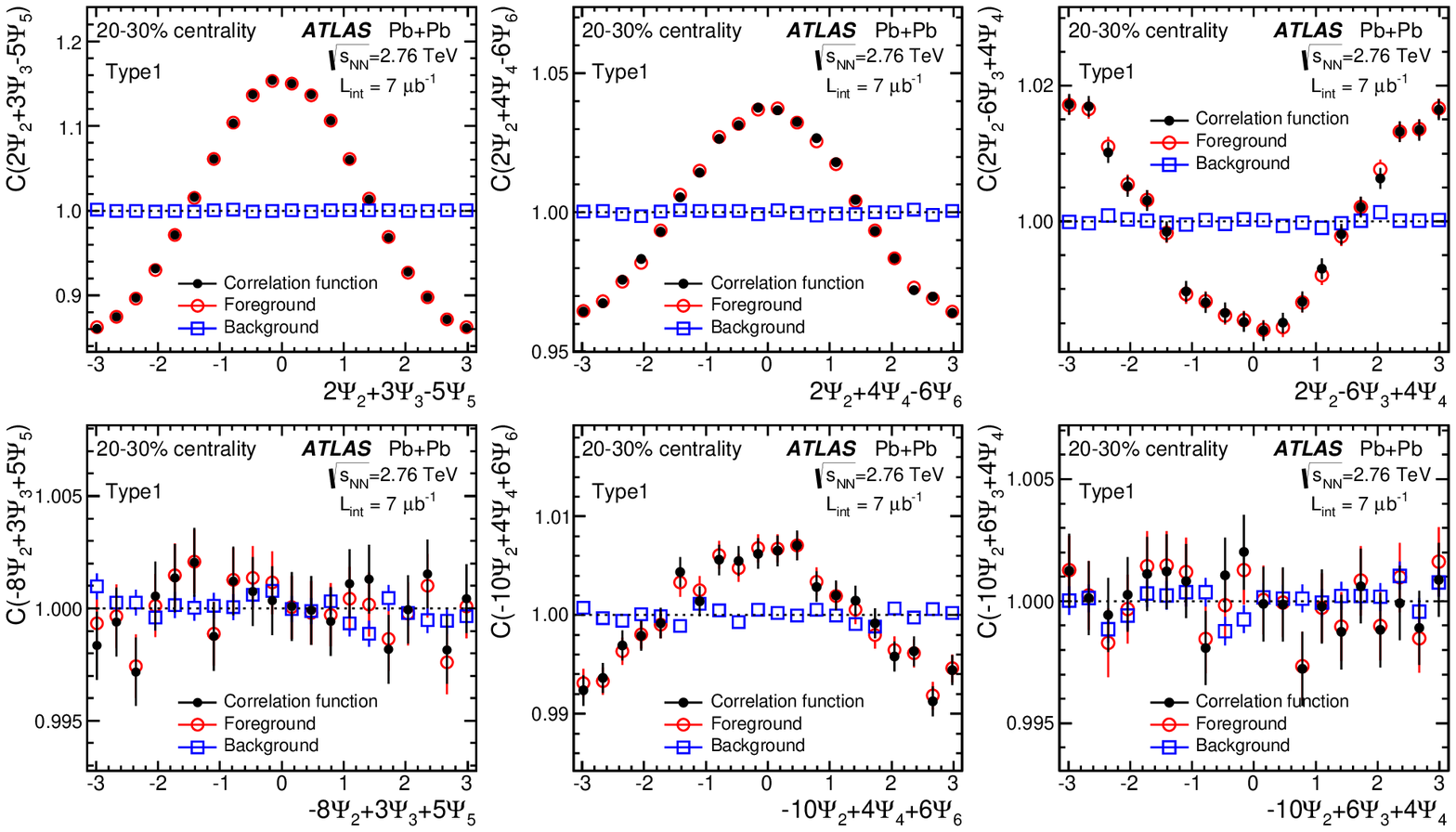}
\end{center}
\caption{\label{fig:3pcraw} (Color online) Relative angle distributions between three event planes from ECal$_{\mathrm{N}}$, FCal and ECal$_{\mathrm{P}}$ defined in Table~\ref{tab:dets} for Type1 correlation in the 20\%--30\% centrality interval for foreground (open circles), background (open squares) and correlation function (filled circles) based on the EP method. The correlation functions give (via equations similar to Eq.~(\ref{eq:2e})) the three-plane correlators defined in Table~\ref{tab:3}. The $y$-axis scales are not the same for all panels.}
\end{figure}

The analysis procedure and the systematic uncertainties discussed above are also valid for the three-plane correlation analysis. However, the 3PC is slightly more complicated because it has three independent measurements for each correlator, which also need to be combined. Figure~\ref{fig:3pcraw} shows the relative angle distributions for various three-plane correlators from the Type1 measurement in the 20\%--30\% centrality interval. The observed correlation signals are calculated as cosine averages of the correlation functions in an obvious generalization of Eq.~(\ref{eq:cf1}):
\begin{eqnarray}
\label{eq:cf2}
C(c_nn\Psi_{n}+c_mm\Psi_{m}+c_hh\Psi_{h}) &=& \frac{S(c_nn\Psi_{n}+c_mm\Psi_{m}+c_hh\Psi_{h})}{B(c_nn\Psi_{n}+c_mm\Psi_{m}+c_hh\Psi_{h})}\;,
\end{eqnarray}
where the background distribution is constructed by requiring that all three angles $\Psi_{n}$, $\Psi_{m}$, and $\Psi_{h}$ are from different events. The correlation functions show significant positive signals for $2\Psi_{2}+3\Psi_{3}-5\Psi_5$,  $2\Psi_{2}+4\Psi_{4}-6\Psi_6$, and $-10\Psi_{2}+4\Psi_{4}+6\Psi_6$, while the signal for $2\Psi_{2}-6\Psi_{3}+4\Psi_4$ is negative, and the signals for the remaining correlators are consistent with zero. 

\begin{figure}[h!]
\begin{center}
\includegraphics[width=1\columnwidth]{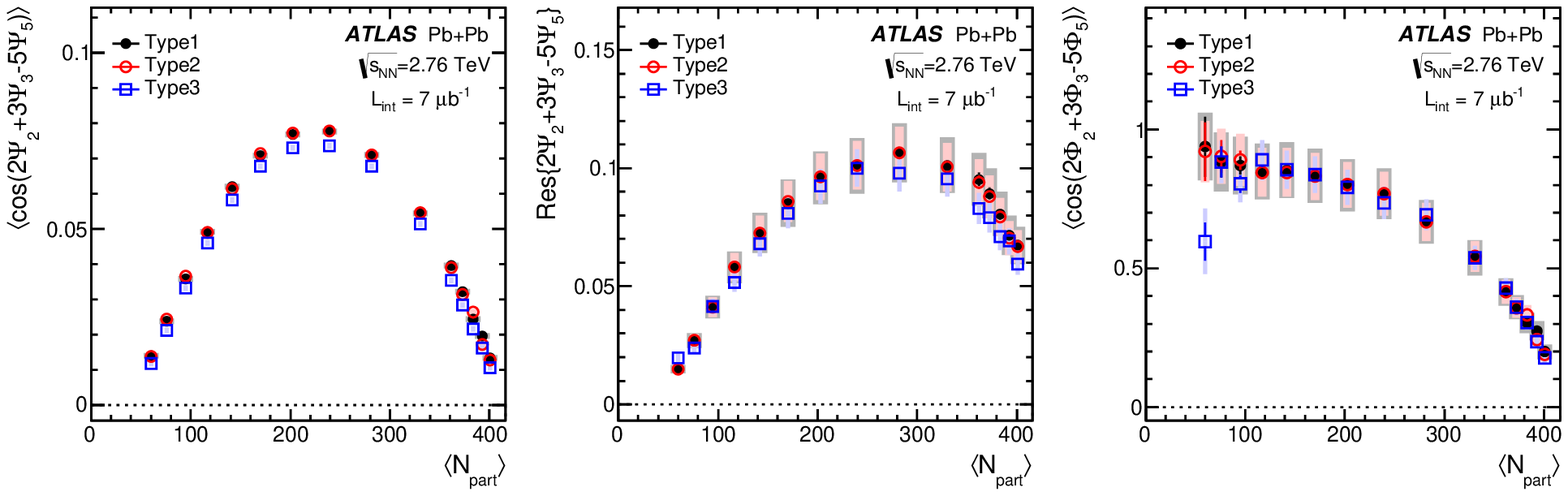}
\end{center}
\caption{\label{fig:3pchain} (Color online) The $\langle N_{\mathrm{part}}\rangle$ dependence of the observed correlation signals (left panel), combined resolutions (middle panel) and corrected signals (right panel) based on the EP method for the three types of event plane combinations for $2\Phi_2+3\Phi_{3}-5\Phi_5$ using the ATLAS calorimeters. The error bars and shaded bands indicate the statistical and systematic uncertainties, respectively.}
\end{figure}

Figure~\ref{fig:3pchain} shows the centrality dependence of the observed correlation signals (left panel), combined resolutions (middle panel) and corrected signals (right panel) for Type1, Type2 and Type3 combinations of $\left\langle\cos (2\Psi_{2}+3\Psi_{3}-5\Psi_5)\right\rangle$. The systematic uncertainty in the observed correlation signals is estimated from the values of $\left\langle \sin \left(c_nn\Psi_n+c_mm\Psi_m+c_hh\Psi_h\right)\right\rangle$, and is calculated by averaging these sine terms over the measured centrality range. This uncertainty is (0.2--1.5)$\times10^{-3}$ in absolute variation, depending on the type of three-plane correlator. The uncertainty in the combined resolution is obtained by propagation from those for the individual resolution factors. Both sources of uncertainties are strongly correlated with centrality, and they are included in the final results (see Sec.~\ref{sec:sys}). Figure~\ref{fig:3pchain} shows that all three types of measurements (Type1, Type2 and Type3) have similar values for the observed signal and the combined resolution. This behavior is expected since the three subevents cover similar rapidity ranges. The three corrected results are statistically combined, and the spreads between them are included in the total systematic uncertainty.

The same analysis procedure is repeated for event-plane correlations obtained via the SP method. The performance of the SP method is found to be very similar to that of the EP method. The magnitudes of the sine terms relative to the cosine terms for both the signal distributions in $k(\Psi_{n}-\Psi_{m})$ and $c_nn\Psi_{n}+c_mm\Psi_{m}+c_hh\Psi_{h}$, as well as the distributions in $n(\Psi_{n}^{\mathrm{A}}-\Psi_{n}^{\mathrm{B}})$ for calculating the resolution factors are found to be nearly the same as those for the EP method. This behavior is quite natural as the effects of detector acceptance are expected to be independent of the strength of the flow signal. The resolution factors and their associated systematic uncertainties are calculated with the same detector combinations as those used for the EP method. The spreads of the results between various detector combinations are included in the systematic uncertainty for the resolution factors. These uncertainties are also found to be strongly correlated between the two methods. The uncorrelated systematic uncertainties between the two methods are evaluated by calculating a double ratio for each detector ``X'' listed in Table~\ref{tab:dets}:
\begin{eqnarray}
\label{eq:res2}
R_{\mathrm{X}}= \frac{\mathrm{Res}\{jn\Psi_{n}\}_{\mathrm{w}}(\mathrm{X,other})/\mathrm{Res}\{jn\Psi_{n}\}_{\mathrm{w}}(\mathrm{X,ref})}{\mathrm{Res}\{jn\Psi_{n}\}(\mathrm{X,other})/\mathrm{Res}\{jn\Psi_{n}\}(\mathrm{X,ref})}\;,
\end{eqnarray}
where the ``ref'' refers to the default detector combination used to calculate the resolution of ``X'', and ``other'' refers to other detector combinations used to evaluate the systematic uncertainties in the resolution of ``X'' via the 2SE and 3SE methods as discussed above (see the paragraph before Eq.~(\ref{eq:cf1b})). The spread of the $R_{\mathrm{X}}$ values provides an estimate of the uncorrelated uncertainty between the two methods for resolution factor $\mathrm{Res}\{jn\Psi_{n}\}$. This uncorrelated uncertainty is typically much smaller than the total systematic uncertainty in the resolution factor in either method.

\subsection{Systematic uncertainties}
\label{sec:sys}
The main systematic uncertainties in the result are introduced and discussed in Sec.~\ref{sec:step} at various key steps of the analysis. This section gives a summary of these uncertainties, and then discusses any additional systematic uncertainties and cross-checks.

The systematic uncertainties associated with the analysis procedure are dominated by contributions from residual detector acceptance effects and uncertainties in the resolution factors. Most detector acceptance effects are expected to cancel in the raw correlation function by dividing the foreground and background distributions (Eqs.~\ref{eq:cf1}--\ref{eq:cf2}). The residual acceptance effects, estimated by the sine terms of the distributions, are found to be (0.2--1.5)$\times10^{-3}$ of the average amplitude of the correlation functions, and are found to be independent of the event centrality. The uncertainties in the resolution factors are calculated from the differences between the 2SE estimate and various 3SE estimates, which are then propagated to give the total uncertainties for the combined resolution factor. These uncertainties are found to be quite similar in the EP and SP methods since they both rely on the same subevent correlations; the larger of the two is quoted as the total systematic uncertainty. The uncorrelated uncertainties are evaluated separately via Eq.~(\ref{eq:res2}), and are used for comparison between the two methods. The uncertainties in the resolution factors are found to depend only weakly on event centrality.

Additional systematic uncertainties include those associated with the trigger and event selections, as well as variations of resolution-corrected signals between different running periods. The former is evaluated by varying the full centrality range by $\pm$ 2\% according to the estimated efficiency of (98$\pm2$)\% for selecting minimum-bias Pb+Pb events. The latter is evaluated by comparing the results obtained independently from three running periods each with 1/3 of the total event statistics. All these uncertainties are generally small, and are quite similar between the EP method and the SP method. Both types of uncertainties are found to be independent of the event centrality.

Tables~\ref{tab:sysfinal2p} and \ref{tab:sysfinal3p} summarize the sources of systematic uncertainties for two-plane and three-plane correlations. The total systematic uncertainties are the quadrature sum of the three sources listed in these tables and the uncertainties associated with residual detector effects. The total uncertainties are found to be nearly independent of the event centrality over the 0\%--50\% centrality range, although a small increase is observed for some of the correlators in the 50\%--75\% centrality range. In most cases, the total systematic uncertainties are dominated by uncertainties associated with the resolution factors. The uncertainties in the resolution correction can become quite sizable when the angles $\Psi_5$ and $\Psi_6$ are involved. This is expected since the higher-order flow signals $v_5$ and $v_6$ are weak, leading to small values of $\mathrm{Res}\{6\Psi_{6}\}$, $\mathrm{Res}\{5\Psi_{5}\}$ and $\mathrm{Res}\{10\Psi_{5}\}$ with large uncertainties.

One important issue in this analysis is the extent to which the measured correlations are biased by short-range correlations such as jet fragmentation, resonance decays and Bose--Einstein correlations. These short-range correlations may contribute to the observed correlation signals and the resolution factors and hence affect the measured correlations. The potential influence of these short-range correlations is studied for the eight two-plane correlators with the EP method. The $\eta$ gap between the two symmetric subevents from ECalFCal, $\eta_{\mathrm{min}}$, is varied in the range of 0 to 8. Seventeen symmetric pairs of subevents are chosen, each corresponding to a different $\eta$ separation. For each case, the observed correlation signals and the resolution factors are obtained using the correlations between these two subevents. Both the observed correlation signals and combined resolutions decrease significantly (by up to a factor of four) as $\eta_{\mathrm{min}}$ is increased. However, the final corrected correlation signals are relatively stable. For example, a gradual change of a few percent is observed for $\eta_{\mathrm{min}}<4$ where the statistical and systematic uncertainties are not very large. This observation strongly suggests that the measurement indeed reflects long-range correlations between the event planes. In most cases, the raw correlation signals decrease smoothly with $\eta_{\mathrm{min}}$. In contrast, the estimated resolution factors have a sharp increase towards small $\eta_{\mathrm{min}}$ in many cases, leading to a suppression of the corrected correlation signals at small $\eta_{\mathrm{min}}$. This behavior suggests that short-range correlations can influence individual harmonics, and hence the resolution factors, but their influences are weak for correlations between EP angles of different order. In all cases, the influences of these short-range correlations are negligible for $\eta_{\mathrm{min}}>0.4$. The choices of the subevents in Table~\ref{tab:dets} have a minimum $\eta$ gap of 0.6, and hence are sufficient to suppress these short-range correlations.

The event-plane correlators measured by the calorimeters are also compared with those obtained independently from the ID for both the EP method and the SP method (see Table~\ref{tab:dets} for the definition of the subevents). Despite the larger fluctuations due to the limited $\eta$ range of its subevents, the results from the ID are consistent with those from the calorimeters (see Appendix). Since the ID is an entirely different type of detector and measures only charged particles, this consistency gives confidence that the measured results are robust. It is argued in Ref.~\cite{Bhalerao:2013ina} that the SP method as defined in Eq.~(\ref{eq:o3ab}) is insensitive to various smearing effects on the weighting factors, such as energy or momentum resolution or multiplicity fluctuations, and as long as these smearings are random and isotropic, they should cancel after averaging over events in the numerator and denominator of Eq.~(\ref{eq:o3ab}). This behavior was checked explicitly in the ID by calculating $\overrightharp{q}_n$ given by Eq.~(\ref{eq:2a}) in several different ways: (1) instead of $u=\pT$ as in the default calculation, the charged particles are set to have equal weight $u=1$,  (2) the weight $u$ is randomly set to be zero for half of the charged particles, or (3) the $\overrightharp{q}_n$ is redefined as $\overrightharp{q}_n\Sigma u_i$ to include explicitly the event-by-event multiplicity fluctuations. The results of all of these cross-checks are consistent with the results of the default calculation.

\begin{table}[!h]
\centering
\small{\begin{tabular}{|l|c|c|c|c|c|c|c|c|}\hline
$\Sigma\Phi$ & $4(\Phi_2-\Phi_{4})$  & $8(\Phi_2-\Phi_{4})$ & $12(\Phi_2-\Phi_{4})$ & $6(\Phi_2-\Phi_{3})$ & $6(\Phi_2-\Phi_{6})$ & $6(\Phi_3-\Phi_{6})$ & $12(\Phi_3-\Phi_{4})$ & $10(\Phi_2-\Phi_{5})$ \tabularnewline\hline
Resolution & 3\%& 7\%& 11\%& 7\%& 10\%& 11\%& 16\%& 9\%\tabularnewline\hline
Trigger \& event sel. & 1--2\%& 1--4\% & 3\%& 3\%& 1--2\%& 1--2\%& $<1\%$ & $<1\%$ \tabularnewline\hline
Run periods      & $<1$\%& 1\% & 5\%& 5\% & 3\% & 3\%& 2\% & 2\% \tabularnewline\hline
\end{tabular}}\normalsize
\caption{\label{tab:sysfinal2p} Three sources of uncertainties for the two-plane correlators, $\left\langle {\cos (\Sigma\Phi) }\right\rangle$, where $\Sigma\Phi=jk(\Phi_n-\Phi_{m})$. They are given as percentage uncertainties.}
\end{table}
\begin{table}[h!]
\centering
\begin{tabular}{|l|c|c|c|}\hline
$\Sigma\Phi$ &$2\Phi_2+3\Phi_{3}-5\Phi_5$ &$2\Phi_2+4\Phi_{4}-6\Phi_6$ &$2\Phi_2-6\Phi_{3}+4\Phi_4$ \tabularnewline\hline 
Resolution         & 10\%& 21\% & 11\% \tabularnewline\hline
Trigger \& event sel. & 1--2\%& 1\% & 3--4\% \tabularnewline\hline
Run periods         & 2\%& 5\% & 5\% \tabularnewline\hline\hline
$\Sigma\Phi$ &$-8\Phi_2+3\Phi_{3}+5\Phi_5$ &$-10\Phi_2+4\Phi_{4}+6\Phi_6$ &$-10\Phi_2+6\Phi_{3}+4\Phi_4$ \tabularnewline\hline 
Resolution & 13\% & 24\% &16\%\tabularnewline\hline
Trigger \& event sel. & 1--3\%& 1--3\% &1--3\%\tabularnewline\hline
Run periods        & 1\%& 5\%  & 5\%\tabularnewline\hline
\end{tabular}
\caption{\label{tab:sysfinal3p} Three sources of uncertainties for the three-plane correlators, $\left\langle {\cos (\Sigma\Phi) }\right\rangle$, where $\Sigma\Phi=c_nn\Phi_{n}+c_mm\Phi_{m}+c_hh\Phi_{h}$. They are given as percentage uncertainties.}
\end{table}

\section{Results and discussions}
\begin{figure}[h!]
\begin{center}
\includegraphics[width=1\columnwidth]{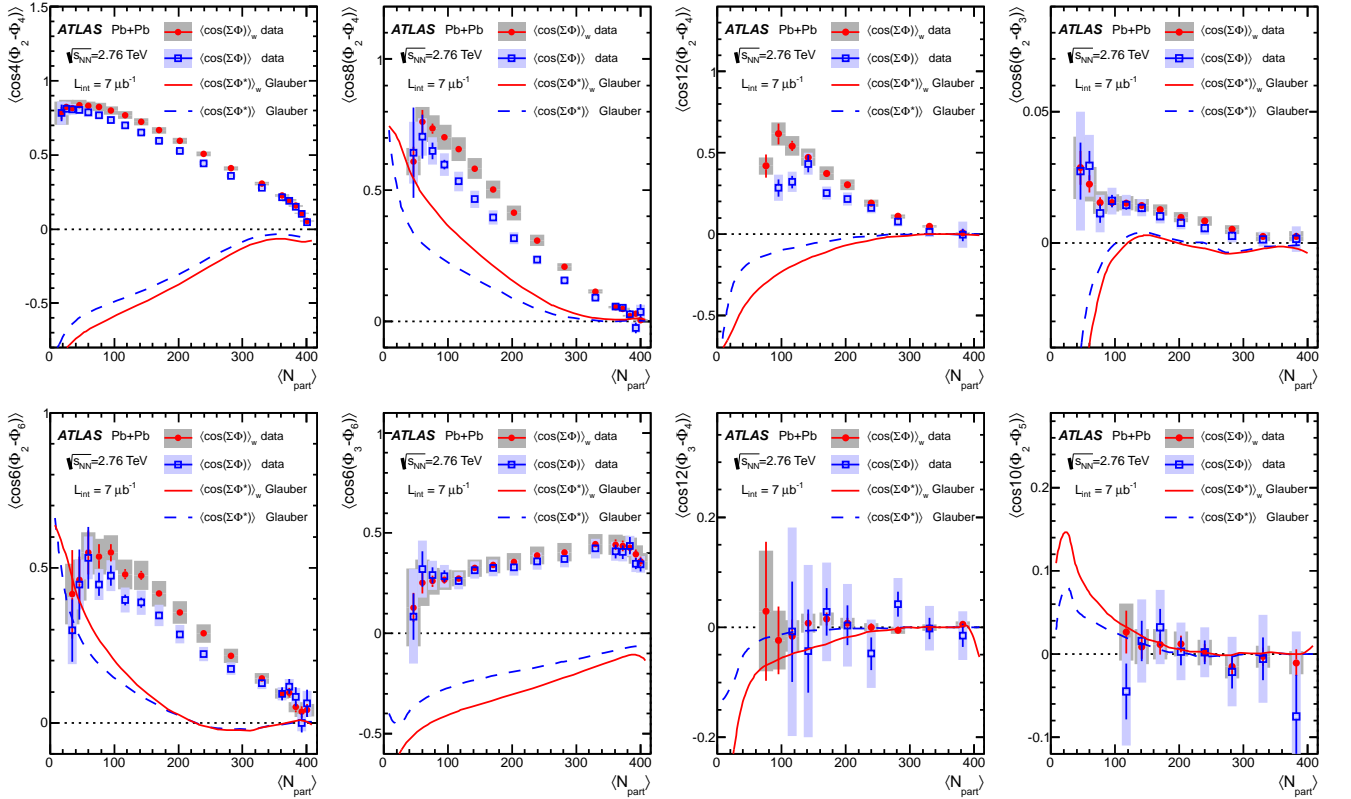}
\end{center}
\caption{\label{fig:result1} (Color online) The centrality dependence of eight two-plane correlators, $\left\langle {\cos (\Sigma\Phi) }\right\rangle$ with $\Sigma\Phi=jk(\Phi_n-\Phi_{m})$ obtained via the SP method (solid symbols) and the EP method (open symbols). The middle two panels in the top row have $j=2$ and $j=3$, respectively, while all other panels have $j=1$. The error bars and shaded bands indicate the statistical uncertainties and total systematic uncertainties, respectively. The expected correlations among participant-plane angles $\Phi^{*}_n$ from a Glauber model are indicated by the solid curves for weighted case (Eq.~(\ref{eq:o3ac})) and dashed lines for the unweighted case.}
\end{figure}

\begin{figure}[h!]
\begin{center}
\includegraphics[width=0.75\columnwidth]{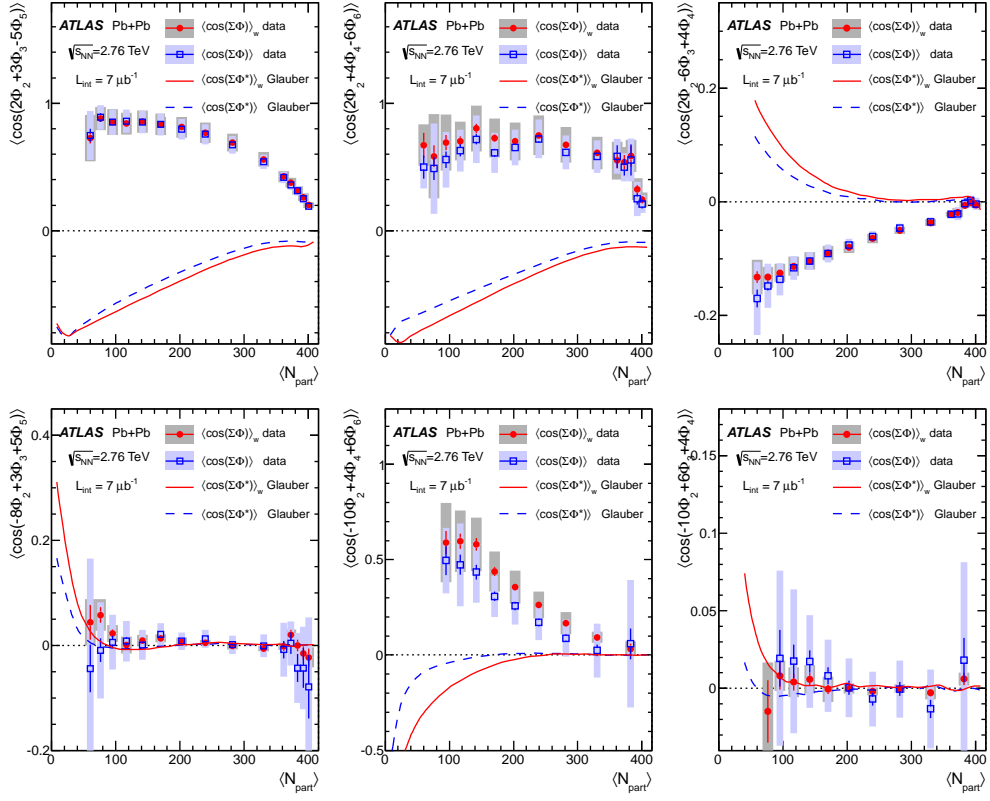}
\end{center}
\caption{\label{fig:result2} (Color online) The centrality dependence of six three-plane correlators, $\left\langle {\cos (\Sigma\Phi) }\right\rangle$ with $\Sigma\Phi=c_nn\Phi_{n}+c_mm\Phi_{m}+c_hh\Phi_{h}$ obtained via the SP method (solid symbols) and the EP method (open symbols). The error bars and shaded bands indicate the statistical uncertainty and total systematic uncertainty, respectively. The expected correlations among participant-plane angles $\Phi^{*}_n$ from a Glauber model are indicated by the solid curves for weighted case (Eq.~(\ref{eq:o3ac})) and dashed lines for the unweighted case.}
\end{figure}

Figures~\ref{fig:result1} and \ref{fig:result2} show the centrality dependence of the two-plane and three-plane correlators, respectively. The results from both the EP method and SP method are shown with their respective systematic uncertainties. These systematic uncertainties are similar in the two methods and are strongly correlated across the centrality range. Strong positive values are observed in most cases and their magnitudes usually decrease with increasing $\langle N_{\mathrm{part}}\rangle$, such as $\left\langle\cos4(\Phi_2-\Phi_4)\right\rangle$, $\left\langle\cos8(\Phi_2-\Phi_4)\right\rangle$, $\left\langle\cos12(\Phi_2-\Phi_4)\right\rangle$, $\left\langle\cos6(\Phi_2-\Phi_6)\right\rangle$, $\left\langle\cos(2\Phi_2+3\Phi_3-5\Phi_5)\right\rangle$, $\left\langle\cos(2\Phi_2+4\Phi_4-6\Phi_6)\right\rangle$ and $\left\langle\cos(-10\Phi_2+4\Phi_4+6\Phi_6)\right\rangle$. The value of $\left\langle\cos6(\Phi_2-\Phi_3)\right\rangle$ is small ($<0.02$), yet exhibits a similar dependence on $\langle N_{\mathrm{part}}\rangle$. A small $\left\langle \cos 6(\Phi_2-\Phi_{3}) \right\rangle$ value in this analysis is a consequence of dividing a small $\left\langle \cos 6(\Psi_2-\Psi_{3}) \right\rangle$ signal (Figure~\ref{fig:raw1}) by a relatively large combined resolution factor (Fig.~\ref{fig:reso1}). Two other correlators show very different trends: the value of $\left\langle\cos6(\Phi_3-\Phi_6)\right\rangle$ increases with $\langle N_{\mathrm{part}}\rangle$, and the value of $\left\langle\cos(2\Phi_2-6\Phi_3+4\Phi_4)\right\rangle$ is negative and its magnitude decreases with $\langle N_{\mathrm{part}}\rangle$. The values of the remaining correlators are consistent with zero. 


\begin{figure}[h!]
\begin{center}
\includegraphics[width=0.85\columnwidth]{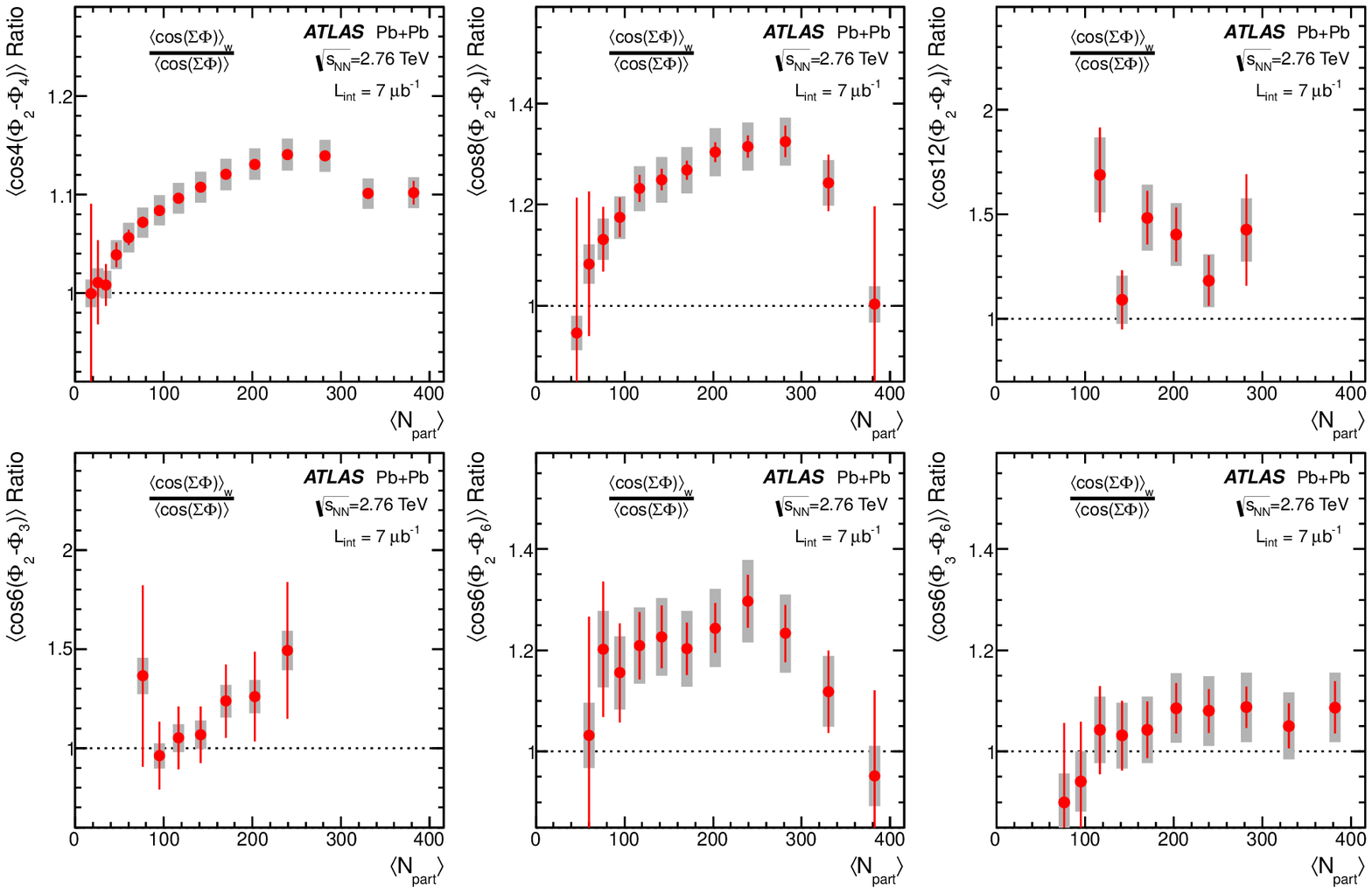}
\end{center}
\caption{\label{fig:rat12} (Color online) The ratios of the SP-method correlators to the EP-method correlators, $\left\langle {\cos (\Sigma\Phi) }\right\rangle_{\mathrm{w}}/\left\langle {\cos (\Sigma\Phi) }\right\rangle$ for several two-plane correlators i.e with $\Sigma\Phi=jk(\Phi_n-\Phi_{m})$. The error bars and shaded bands indicate the statistical uncertainties and total systematic uncertainties, respectively.}
\end{figure}

\begin{figure}[h!]
\begin{center}
\includegraphics[width=1\columnwidth]{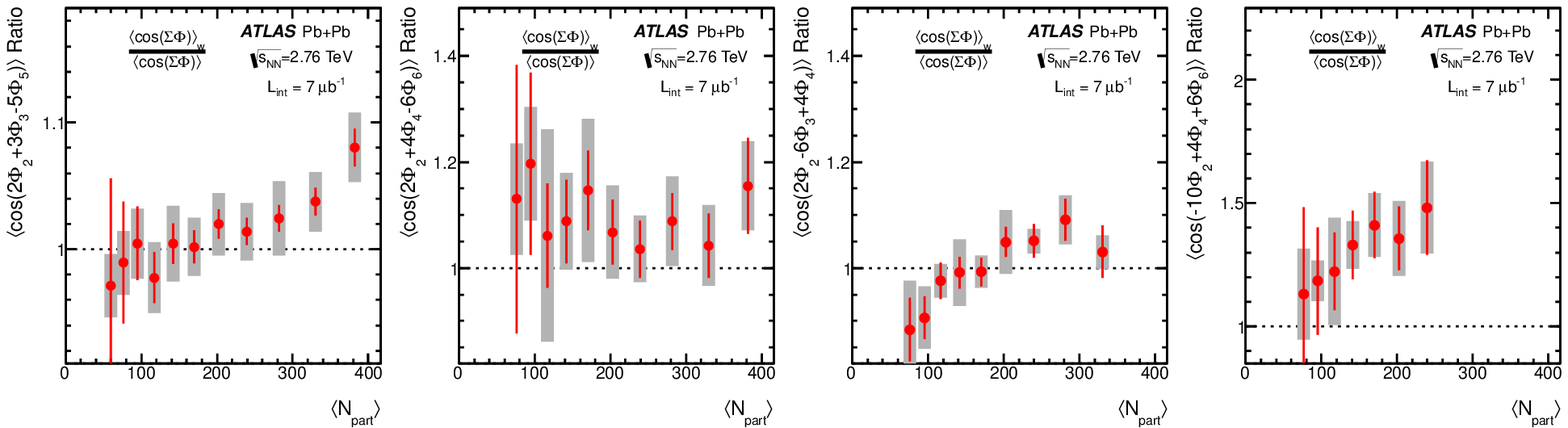}
\end{center}
\caption{\label{fig:rat13} (Color online) The ratios of the SP-method correlators to the EP-method correlators, $\left\langle {\cos (\Sigma\Phi) }\right\rangle_{\mathrm{w}}/\left\langle {\cos (\Sigma\Phi) }\right\rangle$ for several three-plane correlators i.e with  $\Sigma\Phi=c_nn\Phi_{n}+c_mm\Phi_{m}+c_hh\Phi_{h}$. The error bars and shaded bands indicate the statistical uncertainties and total systematic uncertainties, respectively.}
\end{figure}

Figures~\ref{fig:result1} and \ref{fig:result2} also suggest that the magnitude of the correlations from the SP method is always larger than that from the EP method. To better quantify their differences, Figures~\ref{fig:rat12} and \ref{fig:rat13} show the ratio (SP/EP) for some selected two-plane and three-plane correlators, respectively.  As discussed in Sec.~\ref{sec:sys}, the nature of the systematic uncertainties is very similar in the EP and SP methods, and hence these uncertainties mostly cancel in the ratio. The results from the SP method are larger than those from the EP method, and their ratios reach a maximum at around $100<\langle N_{\mathrm{part}}\rangle<300$ range or 10\%--40\% centrality range. The maximum difference is about 10--15\% for most two-plane correlators, but reaches 20--30\% in mid-central collisions for $\left\langle \cos 8(\Phi_2-\Phi_{4}) \right\rangle$ and $\left\langle \cos 6(\Phi_2-\Phi_{6}) \right\rangle$. The differences are smaller for the three-plane correlators, except for $\left\langle\cos(-10\Phi_2+4\Phi_4+6\Phi_6)\right\rangle$.

Figures~\ref{fig:result1} and \ref{fig:result2} also compare the data with the correlators calculated using the participant-plane angles defined in Eq.~(\ref{eq:glau}) from the Glauber model~\cite{Miller:2007ri}. Thirty million events were generated and grouped into centrality intervals according to the impact parameter. If each flow harmonic is driven solely by the corresponding geometric component and the $\Phi_n$ aligns with the $\Phi_n^*$, then the event-plane correlation and participant-plane correlation are expected to have the same sign and show similar centrality dependence. The results in Figs.~\ref{fig:result1} and \ref{fig:result2} show that for several correlators the centrality dependence of the Glauber model predictions show trends similar to the data, although in some cases the sign is opposite, e.g. $\left\langle\cos4(\Phi_2-\Phi_4)\right\rangle$ and $\left\langle\cos(2\Phi_2+3\Phi_3-5\Phi_5)\right\rangle$. In some cases, even the magnitudes of the correlators show opposite centrality dependence between the Glauber model and the data in addition to the sign-flip, such as $\left\langle\cos6(\Phi_3-\Phi_6)\right\rangle$. These discrepancies suggest that in general $\Phi_n$ may not align with $\Phi_n^*$. Indeed, large misalignments between $\Phi_n$ and $\Phi_n^*$ have been observed in event-by-event hydrodynamic model calculations for flow harmonics with $n>3$, and these have been ascribed to the non-linear response of the medium to the fluctuations in the initial geometry~\cite{Qiu:2011iv,Andrade:2010sd}. The non-linear effects are found to be small for lower-order harmonics~\cite{Qiu:2011iv,Niemi:2012aj}, such that $\Phi_n\approx \Phi_n^*$ and $v_n\propto \epsilon_n$ for $n=2$ and 3 or equivalently in the form introduced in Eq.~(\ref{eq:vec}):
\begin{eqnarray}
\label{eq:non0}
v_2e^{i2\Phi_2} \propto \epsilon_2e^{i2\Phi^*_2},\;\; v_3e^{i3\Phi_3} \propto \epsilon_3e^{i3\Phi^*_3}\;.
\end{eqnarray}

Recently, motivated by the preliminary version~\cite{epcorr} of the results presented in this paper, several theory groups calculated the centrality dependence of EP correlators based on hydrodynamic models~\cite{Gardim:2011xv,Teaney:2012ke,Qiu:2012uy,Teaney:2012gu,Teaney:2013dta}. The results of these calculations are in qualitative agreement with the experimental data. The dynamical origin of these correlators has been explained using the so-called single-shot hydrodynamics~\cite{Teaney:2012gu,Teaney:2012ke,Teaney:2013dta}, where small fluctuations are imposed on a smooth average geometry profile, and the hydrodynamic response to these small fluctuations is then derived analytically using a cumulant expansion method. In this analytical approach, the $v_4$ signal comprises a term proportional to the $\epsilon_4$ (linear response term) and a leading non-linear term that is proportional to $\epsilon_2^2$~\cite{Gardim:2011xv,Teaney:2012gu}:
\begin{eqnarray}
\nonumber
v_4e^{i4\Phi_4} &=& \alpha_{_4}\; \epsilon_4e^{i4\Phi^*_4}  + \alpha_{_{2,4}}\; \left(\epsilon_2e^{i2\Phi^*_2}\right)^2 +...\\
               &=& \alpha_{_4}\; \epsilon_4e^{i4\Phi^*_4}  + \beta_{_{2,4}}\; v_2^2e^{i4\Phi_2} +...\;,
\end{eqnarray}
where the second line of the equation is derived from Eq.~(\ref{eq:non0}), and the coefficients $\alpha_4$, $\alpha_{2,4}$ and $\beta_{2,4}$ are all weak functions of centrality. Since $v_2$ increases rapidly for smaller $\langle N_{\mathrm{part}}\rangle$~\cite{Aad:2012bu}, the angle $\Phi_4$ becomes more closely aligned with $\Phi_2$. Hence the centrality dependence of $\left\langle \cos j4(\Phi_2-\Phi_{4}) \right\rangle$ reflects mainly the increase of the $v_2$ as $\langle N_{\mathrm{part}}\rangle$ decreases. 

Similarly, the correlations between $\Phi_2$ and $\Phi_6$ or between $\Phi_3$ and $\Phi_6$ have been explained by the following decomposition of the $v_6$ signal~\cite{Gardim:2011xv,Teaney:2012gu}:
\begin{eqnarray}
\nonumber
v_6e^{i6\Phi_6} &=& \alpha_6 \epsilon_6e^{i6\Phi^*_6} + \alpha_{_{2,6}} \left(\epsilon_2e^{i2\Phi^*_2}\right)^3 + \alpha_{_{3,6}} \left(\epsilon_3e^{i3\Phi^*_3}\right)^2+....\\
               &=& \alpha_6 \epsilon_6e^{i6\Phi^*_6} + \beta_{_{2,6}}  v_2^3e^{i6\Phi_2} + \beta_{_{3,6}} v_3^2e^{i6\Phi_3}+....\;.
\end{eqnarray}
Due to the non-linear contributions, $\Phi_6$ becomes correlated with $\Phi_2$ and $\Phi_3$, even though $\Phi_2$ and $\Phi_3$ are only very weakly correlated. The centrality dependences of $\left\langle \cos 6(\Phi_2-\Phi_{6}) \right\rangle$ and $\left\langle \cos 6(\Phi_3-\Phi_{6}) \right\rangle$ are strongly influenced by the centrality dependence of $v_2$ and $v_3$: since $v_2$ increases for smaller $\langle N_{\mathrm{part}}\rangle$ and $v_3$ is relatively independent of $\langle N_{\mathrm{part}}\rangle$~\cite{Aad:2012bu}, the relative contribution of the second term increases and that of the third term decreases for smaller  $\langle N_{\mathrm{part}}\rangle$, i.e. the collisions become more peripheral. This behavior explains the opposite centrality dependence of $\left\langle \cos 6(\Phi_2-\Phi_{6}) \right\rangle$ and  $\left\langle \cos 6(\Phi_3-\Phi_{6}) \right\rangle$.

In the same manner, the correlation between $\Phi_2, \Phi_3$ and $\Phi_5$ has been explained by the following decomposition of the $v_5$ signal~\cite{Gardim:2011xv,Teaney:2012gu}:
\begin{eqnarray}
\nonumber
v_5e^{i5\Phi_5} &=& \alpha_5 \epsilon_5e^{i5\Phi^*_5} + \alpha_{_{2,3,5}} \epsilon_2e^{i2\Phi^*_2}\epsilon_3e^{i3\Phi^*_3}+....\\
               &=& \alpha_5 \epsilon_5e^{i5\Phi^*_5} + \beta_{_{2,3,5}} v_2v_3e^{i(2\Phi_2+3\Phi_3)}+...\;.
\end{eqnarray}
The coupling between $v_5$ and $v_2v_3$ explains qualitatively the centrality dependence of the correlator $\left\langle\cos(2\Phi_2+3\Phi_3-5\Phi_5)\right\rangle$.

\begin{figure}[h!]
\begin{center}
\includegraphics[width=0.8\columnwidth]{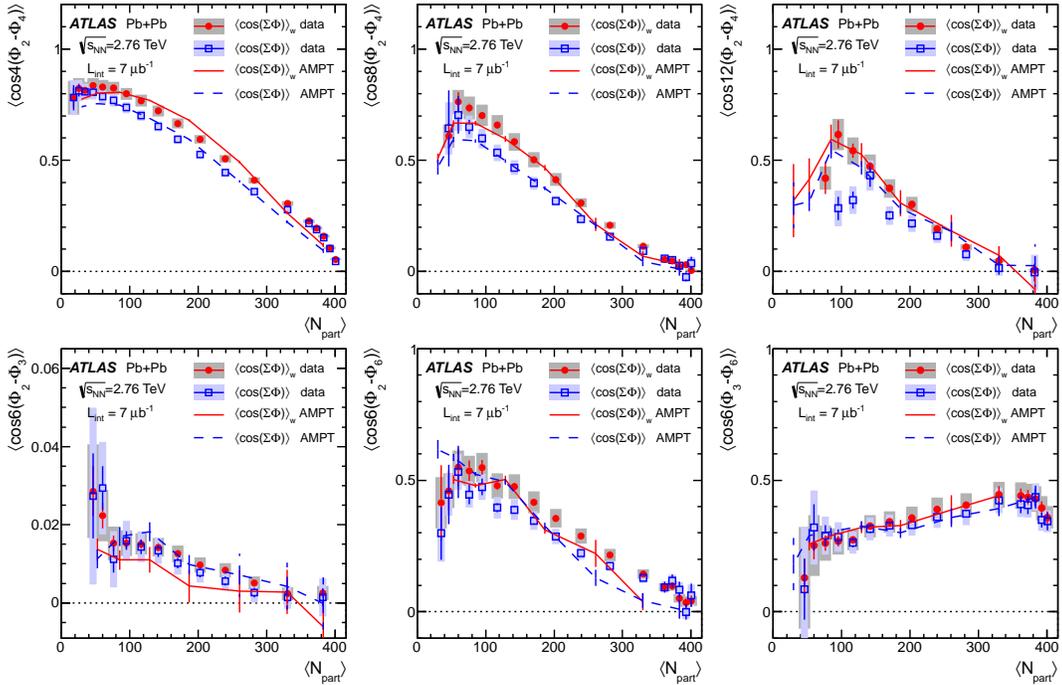}
\end{center}
\caption{\label{fig:comp2pcampt} (Color online) Comparison of six two-plane correlators, $\left\langle {\cos (\Sigma\Phi) }\right\rangle$ with $\Sigma\Phi=jk(\Phi_n-\Phi_{m})$, with results from the AMPT model calculated via the SP method (solid lines) and the EP method (dashed lines) from Ref.~\cite{Bhalerao:2013ina}. The error bars on the lines represent the statistical uncertainties in the calculation.}
\end{figure}

\begin{figure}[h!]
\begin{center}
\includegraphics[width=1\columnwidth]{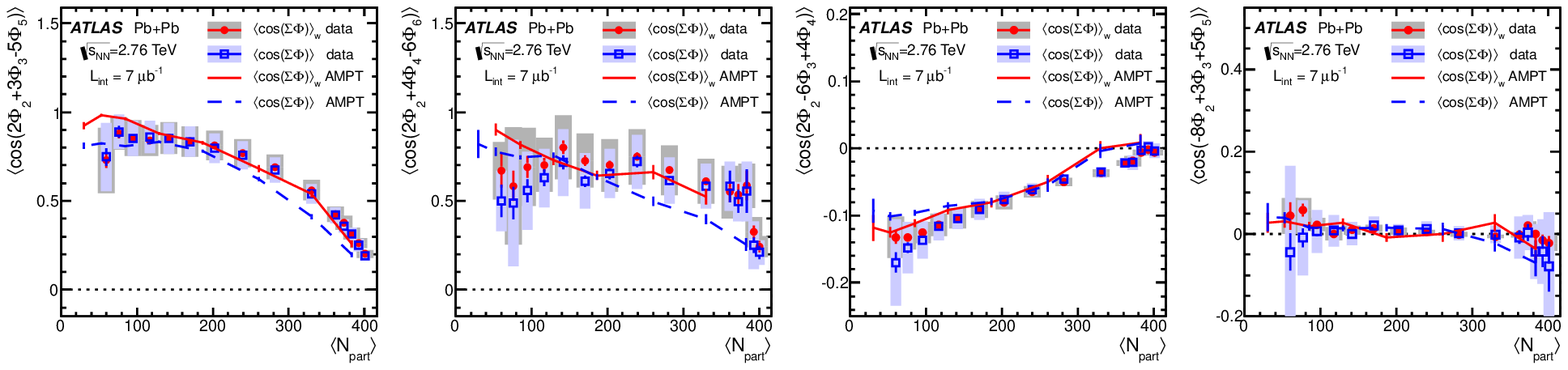}
\end{center}
\caption{\label{fig:comp3pcampt} (Color online) Comparison of four three-plane correlators, $\left\langle {\cos (\Sigma\Phi) }\right\rangle$ with $\Sigma\Phi=c_nn\Phi_{n}+c_mm\Phi_{m}+c_hh\Phi_{h}$, with results from the AMPT model calculated via the SP method (solid lines) and the EP method (dashed lines) from Ref.~\cite{Bhalerao:2013ina}. The error bars on the curves represent the statistical uncertainties in the calculation.}
\end{figure}

A multi-phase transport (AMPT) model~\cite{Lin:2004en} is frequently used to study the harmonic flow coefficients $v_n$ and to study the relation of $v_n$ to the initial geometry. The AMPT model combines the initial-state geometry fluctuations of the Glauber model and final-state interactions through a parton and hadron transport model. The AMPT model generates collective flow by elastic scatterings in the partonic and hadronic phase and was shown to reproduce the $v_n$ values~\cite{Xu:2011jm} and the particle multiplicity~\cite{Xu:2011fi} reasonably well. As a full event generator, the AMPT model allows the generated events to be analyzed with the same procedures as in the data. Figures~\ref{fig:comp2pcampt} and \ref{fig:comp3pcampt} compare some selected correlators (six two-plane correlators and four three-plane correlators) with a prediction~\cite{Bhalerao:2013ina} from the AMPT model. Good agreement is observed between the data and the calculation, and in particular the model predicts correctly the stronger signal observed with the SP method.


\section{Conclusions}
Measurements of fourteen correlators between two and three event planes, $\left\langle {\cos jk(\Phi_n-\Phi_{m}) } \right\rangle$ and $\left\langle \cos \left(c_nn\Psi_n+c_mm\Psi_m+c_hh\Psi_h\right)\right\rangle$, respectively, are presented using 7 $\mu$b$^{-1}$ of Pb+Pb collision data at $\sqrt{s_{_{\mathrm{NN}}}}=2.76$~TeV collected by the ATLAS experiment at the LHC. These correlations are estimated from correlations of observed event-plane angles measured in the calorimeters over a large pseudorapidity range $|\eta|<4.8$ using both a standard event-plane method and a scalar-product method. Significant positive correlation signals are observed for $4(\Phi_2-\Phi_4)$, $6(\Phi_2-\Phi_6)$, $6(\Phi_3-\Phi_6)$, $2\Phi_2+3\Phi_3-5\Phi_5$, $2\Phi_2+4\Phi_4-6\Phi_6$ and $-10\Phi_2+4\Phi_4+6\Phi_6$. The correlation signals are negative for $2\Phi_2-6\Phi_3+4\Phi_4$. The magnitudes of the correlations from the scalar-product method are observed to be systematically larger than those obtained from the event-plane method. The centrality dependence of most correlators is found to be very different from that predicted by a Glauber model. However, calculations based on the same Glauber model, but including the final-state collective dynamics, are able to describe qualitatively, and in many cases also quantitatively, the centrality dependence of the measured correlators. These observations suggest that both the fluctuations in the initial geometry and non-linear mixing between different harmonics in the final state are important for creating these correlations in momentum space. A detailed theoretical description of these correlations can improve our present understanding of the space-time evolution of the hot and dense matter created in heavy-ion collisions.

% Acknowledgements for papers with collision data
% Version 19-Feb-2014
\section*{Acknowledgments}
% Standard acknowledgements start here
%----------------------------------------------
We thank CERN for the very successful operation of the LHC, as well as the
support staff from our institutions without whom ATLAS could not be
operated efficiently.

We acknowledge the support of ANPCyT, Argentina; YerPhI, Armenia; ARC,
Australia; BMWF and FWF, Austria; ANAS, Azerbaijan; SSTC, Belarus; CNPq and FAPESP,
Brazil; NSERC, NRC and CFI, Canada; CERN; CONICYT, Chile; CAS, MOST and NSFC,
China; COLCIENCIAS, Colombia; MSMT CR, MPO CR and VSC CR, Czech Republic;
DNRF, DNSRC and Lundbeck Foundation, Denmark; EPLANET, ERC and NSRF, European Union;
IN2P3-CNRS, CEA-DSM/IRFU, France; GNSF, Georgia; BMBF, DFG, HGF, MPG and AvH
Foundation, Germany; GSRT and NSRF, Greece; ISF, MINERVA, GIF, I-CORE and Benoziyo Center,
Israel; INFN, Italy; MEXT and JSPS, Japan; CNRST, Morocco; FOM and NWO,
Netherlands; BRF and RCN, Norway; MNiSW and NCN, Poland; GRICES and FCT, Portugal; MNE/IFA, Romania; MES of Russia and ROSATOM, Russian Federation; JINR; MSTD,
Serbia; MSSR, Slovakia; ARRS and MIZ\v{S}, Slovenia; DST/NRF, South Africa;
MINECO, Spain; SRC and Wallenberg Foundation, Sweden; SER, SNSF and Cantons of
Bern and Geneva, Switzerland; NSC, Taiwan; TAEK, Turkey; STFC, the Royal
Society and Leverhulme Trust, United Kingdom; DOE and NSF, United States of
America.

The crucial computing support from all WLCG partners is acknowledged
gratefully, in particular from CERN and the ATLAS Tier-1 facilities at
TRIUMF (Canada), NDGF (Denmark, Norway, Sweden), CC-IN2P3 (France),
KIT/GridKA (Germany), INFN-CNAF (Italy), NL-T1 (Netherlands), PIC (Spain),
ASGC (Taiwan), RAL (UK) and BNL (USA) and in the Tier-2 facilities
worldwide.
%----------------------------------------------

%\bibliography{epcorr_v4}{}

\begin{thebibliography}{47}%
\makeatletter
\providecommand \@ifxundefined [1]{%
 \@ifx{#1\undefined}
}%
\providecommand \@ifnum [1]{%
 \ifnum #1\expandafter \@firstoftwo
 \else \expandafter \@secondoftwo
 \fi
}%
\providecommand \@ifx [1]{%
 \ifx #1\expandafter \@firstoftwo
 \else \expandafter \@secondoftwo
 \fi
}%
\providecommand \natexlab [1]{#1}%
\providecommand \enquote  [1]{``#1''}%
\providecommand \bibnamefont  [1]{#1}%
\providecommand \bibfnamefont [1]{#1}%
\providecommand \citenamefont [1]{#1}%
\providecommand \href@noop [0]{\@secondoftwo}%
\providecommand \href [0]{\begingroup \@sanitize@url \@href}%
\providecommand \@href[1]{\@@startlink{#1}\@@href}%
\providecommand \@@href[1]{\endgroup#1\@@endlink}%
\providecommand \@sanitize@url [0]{\catcode `\\12\catcode `\$12\catcode
  `\&12\catcode `\#12\catcode `\^12\catcode `\_12\catcode `\%12\relax}%
\providecommand \@@startlink[1]{}%
\providecommand \@@endlink[0]{}%
\providecommand \url  [0]{\begingroup\@sanitize@url \@url }%
\providecommand \@url [1]{\endgroup\@href {#1}{\urlprefix }}%
\providecommand \urlprefix  [0]{URL }%
\providecommand \Eprint [0]{\href }%
\providecommand \doibase [0]{http://dx.doi.org/}%
\providecommand \selectlanguage [0]{\@gobble}%
\providecommand \bibinfo  [0]{\@secondoftwo}%
\providecommand \bibfield  [0]{\@secondoftwo}%
\providecommand \translation [1]{[#1]}%
\providecommand \BibitemOpen [0]{}%
\providecommand \bibitemStop [0]{}%
\providecommand \bibitemNoStop [0]{.\EOS\space}%
\providecommand \EOS [0]{\spacefactor3000\relax}%
\providecommand \BibitemShut  [1]{\csname bibitem#1\endcsname}%
\let\auto@bib@innerbib\@empty
%</preamble>
\bibitem [{\citenamefont {Voloshin}\ \emph {et~al.}(2008)\citenamefont
  {Voloshin}, \citenamefont {Poskanzer},\ and\ \citenamefont
  {Snellings}}]{Voloshin:2008dg}%
  \BibitemOpen
  \bibfield  {author} {\bibinfo {author} {\bibfnamefont {S.~A.}\ \bibnamefont
  {Voloshin}}, \bibinfo {author} {\bibfnamefont {A.~M.}\ \bibnamefont
  {Poskanzer}}, \ and\ \bibinfo {author} {\bibfnamefont {R.}~\bibnamefont
  {Snellings}},\ }\ \Eprint
  {http://arxiv.org/abs/0809.2949} {arXiv:0809.2949 [nucl-ex]} \BibitemShut
  {NoStop}%
%%CITATION = ARXIV:0809.2949;%%
\bibitem [{\citenamefont {Heinz}\ and\ \citenamefont
  {Snellings}(2013)}]{Heinz:2013th}%
  \BibitemOpen
  \bibfield  {author} {\bibinfo {author} {\bibfnamefont {U.}~\bibnamefont
  {Heinz}}\ and\ \bibinfo {author} {\bibfnamefont {R.}~\bibnamefont
  {Snellings}},\ }\href {\doibase 10.1146/annurev-nucl-102212-170540}
  {\bibfield  {journal} {\bibinfo  {journal} {Ann.~Rev.~Nucl.~Part.~Sci.}\ }\textbf
  {\bibinfo {volume} {63}},\ \bibinfo {pages} {123} (\bibinfo {year} {2013})},\
  \Eprint {http://arxiv.org/abs/1301.2826} {arXiv:1301.2826 [nucl-th]}
  \BibitemShut {NoStop}%
%%CITATION = ARXIV:1301.2826;%%
\bibitem [{\citenamefont {Teaney}(2009)}]{Teaney:2009qa}%
  \BibitemOpen
  \bibfield  {author} {\bibinfo {author} {\bibfnamefont {D.~A.}\ \bibnamefont
  {Teaney}}},\ \Eprint
  {http://arxiv.org/abs/0905.2433} {arXiv:0905.2433 [nucl-th]} \BibitemShut
  {NoStop}%
%%CITATION = ARXIV:0905.2433;%%
\bibitem [{\citenamefont {Poskanzer}\ and\ \citenamefont
  {Voloshin}(1998)}]{Poskanzer:1998yz}%
  \BibitemOpen
  \bibfield  {author} {\bibinfo {author} {\bibfnamefont {A.~M.}\ \bibnamefont
  {Poskanzer}}\ and\ \bibinfo {author} {\bibfnamefont {S.}~\bibnamefont
  {Voloshin}},\ }\href {\doibase 10.1103/PhysRevC.58.1671} {\bibfield
  {journal} {\bibinfo  {journal} {Phys.~Rev.}\ }\textbf {\bibinfo {volume}
  {C~58}},\ \bibinfo {pages} {1671} (\bibinfo {year} {1998})},\ \Eprint
  {http://arxiv.org/abs/nucl-ex/9805001} {arXiv:nucl-ex/9805001 [nucl-ex]}
  \BibitemShut {NoStop}%
%%CITATION = NUCL-EX/9805001;%%
\bibitem [{\citenamefont {Gardim}\ \emph {et~al.}(2012)\citenamefont {Gardim},
  \citenamefont {Grassi}, \citenamefont {Luzum},\ and\ \citenamefont
  {Ollitrault}}]{Gardim:2011xv}%
  \BibitemOpen
  \bibfield  {author} {\bibinfo {author} {\bibfnamefont {F.~G.}\ \bibnamefont
  {Gardim}}, \bibinfo {author} {\bibfnamefont {F.}~\bibnamefont {Grassi}},
  \bibinfo {author} {\bibfnamefont {M.}~\bibnamefont {Luzum}}, \ and\ \bibinfo
  {author} {\bibfnamefont {J.-Y.}\ \bibnamefont {Ollitrault}},\ }\href
  {\doibase 10.1103/PhysRevC.85.024908} {\bibfield  {journal} {\bibinfo
  {journal} {Phys.~Rev.}\ }\textbf {\bibinfo {volume} {C~85}},\ \bibinfo {pages}
  {024908} (\bibinfo {year} {2012})},\ \Eprint {http://arxiv.org/abs/1111.6538}
  {arXiv:1111.6538 [nucl-th]} \BibitemShut {NoStop}%
%%CITATION = ARXIV:1111.6538;%%
\bibitem [{\citenamefont {Aamodt}\ \emph {et~al.}(2012)\citenamefont {Aamodt}
  \emph {et~al.}}]{Aamodt:2011by}%
  \BibitemOpen
  \bibfield  {author} {\bibinfo {author} {\bibfnamefont {ALICE Collaboration,~}}}\href
  {\doibase 10.1016/j.physletb.2012.01.060} {\bibfield  {journal} {\bibinfo
  {journal} {Phys.~Lett.}\ }\textbf {\bibinfo {volume} {B~708}},\ \bibinfo
  {pages} {249} (\bibinfo {year} {2012})},\ \Eprint
  {http://arxiv.org/abs/1109.2501} {arXiv:1109.2501 [nucl-ex]} \BibitemShut
  {NoStop}%
%%CITATION = ARXIV:1109.2501;%%
\bibitem [{\citenamefont {Chatrchyan}\ \emph {et~al.}(2012)\citenamefont
  {Chatrchyan} \emph {et~al.}}]{CMS:2012wg}%
  \BibitemOpen
  \bibfield  {author} {\bibinfo {author} {\bibfnamefont  {CMS Collaboration,~}}}\href
  {\doibase 10.1140/epjc/s10052-012-2012-3} {\bibfield  {journal} {\bibinfo
  {journal} {Eur.~Phys.~J.}\ }\textbf {\bibinfo {volume} {C~72}},\ \bibinfo
  {pages} {2012} (\bibinfo {year} {2012})},\ \Eprint
  {http://arxiv.org/abs/1201.3158} {arXiv:1201.3158 [nucl-ex]} \BibitemShut
  {NoStop}%
%%CITATION = ARXIV:1201.3158;%%
\bibitem [{\citenamefont {Aad}\ \emph {et~al.}(2012{\natexlab{a}})\citenamefont
  {Aad} \emph {et~al.}}]{Aad:2012bu}%
  \BibitemOpen
  \bibfield  {author} {\bibinfo {author} {\bibfnamefont {ATLAS Collaboration,~}} }\href {\doibase
  10.1103/PhysRevC.86.014907} {\bibfield  {journal} {\bibinfo  {journal}
  {Phys.~Rev.}\ }\textbf {\bibinfo {volume} {C~86}},\ \bibinfo {pages} {014907}
  (\bibinfo {year} {2012}{\natexlab{a}})},\ \Eprint
  {http://arxiv.org/abs/1203.3087} {arXiv:1203.3087 [hep-ex]} \BibitemShut
  {NoStop}%
%%CITATION = ARXIV:1203.3087;%%
\bibitem [{\citenamefont {Miller}\ \emph {et~al.}(2007)\citenamefont {Miller},
  \citenamefont {Reygers}, \citenamefont {Sanders},\ and\ \citenamefont
  {Steinberg}}]{Miller:2007ri}%
  \BibitemOpen
  \bibfield  {author} {\bibinfo {author} {\bibfnamefont {M.~L.}\ \bibnamefont
  {Miller}}, \bibinfo {author} {\bibfnamefont {K.}~\bibnamefont {Reygers}},
  \bibinfo {author} {\bibfnamefont {S.~J.}\ \bibnamefont {Sanders}}, \ and\
  \bibinfo {author} {\bibfnamefont {P.}~\bibnamefont {Steinberg}},\ }\href
  {\doibase 10.1146/annurev.nucl.57.090506.123020} {\bibfield  {journal}
  {\bibinfo  {journal} {Ann.~Rev.~Nucl.~Part.~Sci.}\ }\textbf {\bibinfo {volume}
  {57}},\ \bibinfo {pages} {205} (\bibinfo {year} {2007})},\ \Eprint
  {http://arxiv.org/abs/nucl-ex/0701025} {arXiv:nucl-ex/0701025 [nucl-ex]}
  \BibitemShut {NoStop}%
%%CITATION = NUCL-EX/0701025;%%
\bibitem [{\citenamefont {Teaney}\ and\ \citenamefont
  {Yan}(2011)}]{Teaney:2010vd}%
  \BibitemOpen
  \bibfield  {author} {\bibinfo {author} {\bibfnamefont {D.}~\bibnamefont
  {Teaney}}\ and\ \bibinfo {author} {\bibfnamefont {L.}~\bibnamefont {Yan}},\
  }\href {\doibase 10.1103/PhysRevC.83.064904} {\bibfield  {journal} {\bibinfo
  {journal} {Phys.~Rev.}\ }\textbf {\bibinfo {volume} {C~83}},\ \bibinfo {pages}
  {064904} (\bibinfo {year} {2011})},\ \Eprint {http://arxiv.org/abs/1010.1876}
  {arXiv:1010.1876 [nucl-th]} \BibitemShut {NoStop}%
%%CITATION = ARXIV:1010.1876;%%
\bibitem [{\citenamefont {Qiu}\ and\ \citenamefont {Heinz}(2011)}]{Qiu:2011iv}%
  \BibitemOpen
  \bibfield  {author} {\bibinfo {author} {\bibfnamefont {Z.}~\bibnamefont
  {Qiu}}\ and\ \bibinfo {author} {\bibfnamefont {U.~W.}\ \bibnamefont
  {Heinz}},\ }\href {\doibase 10.1103/PhysRevC.84.024911} {\bibfield  {journal}
  {\bibinfo  {journal} {Phys.~Rev.}\ }\textbf {\bibinfo {volume} {C~84}},\
  \bibinfo {pages} {024911} (\bibinfo {year} {2011})},\ \Eprint
  {http://arxiv.org/abs/1104.0650} {arXiv:1104.0650 [nucl-th]} \BibitemShut
  {NoStop}%
%%CITATION = ARXIV:1104.0650;%%
\bibitem [{\citenamefont {Adare}\ \emph {et~al.}(2011)\citenamefont {Adare}
  \emph {et~al.}}]{Adare:2011tg}%
  \BibitemOpen
  \bibfield  {author} {\bibinfo {author} {\bibfnamefont {A.}~\bibnamefont
  {Adare}} \emph {et~al.} (\bibinfo {collaboration} {PHENIX Collaboration}),\
  }\href {\doibase 10.1103/PhysRevLett.107.252301} {\bibfield  {journal}
  {\bibinfo  {journal} {Phys.~Rev.~Lett.}\ }\textbf {\bibinfo {volume} {107}},\
  \bibinfo {pages} {252301} (\bibinfo {year} {2011})},\ \Eprint
  {http://arxiv.org/abs/1105.3928} {arXiv:1105.3928 [nucl-ex]} \BibitemShut
  {NoStop}%
%%CITATION = ARXIV:1105.3928;%%
\bibitem [{\citenamefont {Adamczyk}\ \emph {et~al.}(2013)\citenamefont
  {Adamczyk} \emph {et~al.}}]{Adamczyk:2013waa}%
  \BibitemOpen
  \bibfield  {author} {\bibinfo {author} {\bibfnamefont {L.}~\bibnamefont
  {Adamczyk}} \emph {et~al.} (\bibinfo {collaboration} {STAR Collaboration}),\
  }\href {\doibase 10.1103/PhysRevC.88.014904} {\bibfield  {journal} {\bibinfo
  {journal} {Phys.~Rev.}\ }\textbf {\bibinfo {volume} {C~88}},\ \bibinfo {pages}
  {014904} (\bibinfo {year} {2013})},\ \Eprint {http://arxiv.org/abs/1301.2187}
  {arXiv:1301.2187 [nucl-ex]} \BibitemShut {NoStop}%
%%CITATION = ARXIV:1301.2187;%%
\bibitem [{\citenamefont {Aamodt}\ \emph {et~al.}(2010)\citenamefont {Aamodt}
  \emph {et~al.}}]{Aamodt:2010pa}%
  \BibitemOpen
  \bibfield  {author} {\bibinfo {author} {\bibfnamefont {ALICE Collaboration,~}}
  }\href {\doibase 10.1103/PhysRevLett.105.252302} {\bibfield  {journal}
  {\bibinfo  {journal} {Phys.~Rev.~Lett.}\ }\textbf {\bibinfo {volume} {105}},\
  \bibinfo {pages} {252302} (\bibinfo {year} {2010})},\ \Eprint
  {http://arxiv.org/abs/1011.3914} {arXiv:1011.3914 [nucl-ex]} \BibitemShut
  {NoStop}%
%%CITATION = ARXIV:1011.3914;%%
\bibitem [{\citenamefont {Aad}\ \emph {et~al.}(2013{\natexlab{a}})\citenamefont
  {Aad} \emph {et~al.}}]{Aad:2013xma}%
  \BibitemOpen
  \bibfield  {author} {\bibinfo {author} {\bibfnamefont {ATLAS Collaboration,~}}
  }\href {\doibase 10.1007/JHEP11(2013)183} {\bibfield  {journal} {\bibinfo
  {journal} {J.~High~Energy~Phy.}\ }\textbf {\bibinfo {volume} {1311}},\ \bibinfo {pages}
  {183} (\bibinfo {year} {2013}{\natexlab{a}})},\ \Eprint
  {http://arxiv.org/abs/1305.2942} {arXiv:1305.2942 [hep-ex]} \BibitemShut
  {NoStop}%
%%CITATION = ARXIV:1305.2942;%%
\bibitem [{\citenamefont {Chatrchyan}\ \emph {et~al.}(2013)\citenamefont
  {Chatrchyan} \emph {et~al.}}]{Chatrchyan:2013kba}%
  \BibitemOpen
  \bibfield  {author} {\bibinfo {author} {\bibfnamefont {CMS Collaboration,~}}
  }\href {\doibase 10.1103/PhysRevC.89.044906} {\bibfield
  {journal} {\bibinfo  {journal} {Phys.~Rev.}\ }\textbf {\bibinfo {volume}
  {C~89}},\ \bibinfo {pages} {044906} (\bibinfo {year} {2014})},\ \Eprint
  {http://arxiv.org/abs/1310.8651} {arXiv:1310.8651 [nucl-ex]} \BibitemShut
  {NoStop}%
%%CITATION = ARXIV:1310.8651;%%
\bibitem [{\citenamefont {Nagle}\ and\ \citenamefont
  {McCumber}(2011)}]{Nagle:2010zk}%
  \BibitemOpen
  \bibfield  {author} {\bibinfo {author} {\bibfnamefont {J.~L.}\ \bibnamefont
  {Nagle}}\ and\ \bibinfo {author} {\bibfnamefont {M.~P.}\ \bibnamefont
  {McCumber}},\ }\href {\doibase 10.1103/PhysRevC.83.044908} {\bibfield
  {journal} {\bibinfo  {journal} {Phys.~Rev.}\ }\textbf {\bibinfo {volume}
  {C~83}},\ \bibinfo {pages} {044908} (\bibinfo {year} {2011})},\ \Eprint
  {http://arxiv.org/abs/1011.1853} {arXiv:1011.1853 [nucl-ex]} \BibitemShut
  {NoStop}%
%%CITATION = ARXIV:1011.1853;%%
\bibitem [{\citenamefont {Adare}\ \emph {et~al.}(2010)\citenamefont {Adare}
  \emph {et~al.}}]{Adare:2010ux}%
  \BibitemOpen
  \bibfield  {author} {\bibinfo {author} {\bibfnamefont {A.}~\bibnamefont
  {Adare}} \emph {et~al.} (\bibinfo {collaboration} {PHENIX Collaboration}),\
  }\href {\doibase 10.1103/PhysRevLett.105.062301} {\bibfield  {journal}
  {\bibinfo  {journal} {Phys.~Rev.~Lett.}\ }\textbf {\bibinfo {volume} {105}},\
  \bibinfo {pages} {062301} (\bibinfo {year} {2010})},\ \Eprint
  {http://arxiv.org/abs/1003.5586} {arXiv:1003.5586 [nucl-ex]} \BibitemShut
  {NoStop}%
%%CITATION = ARXIV:1003.5586;%%
\bibitem [{\citenamefont {Aamodt}\ \emph {et~al.}(2011)\citenamefont {Aamodt}
  \emph {et~al.}}]{ALICE:2011ab}%
  \BibitemOpen
  \bibfield  {author} {\bibinfo {author} {\bibfnamefont {ALICE Collaboration,~}}}\href
  {\doibase 10.1103/PhysRevLett.107.032301} {\bibfield  {journal} {\bibinfo
  {journal} {Phys.~Rev.~Lett.}\ }\textbf {\bibinfo {volume} {107}},\ \bibinfo
  {pages} {032301} (\bibinfo {year} {2011})},\ \Eprint
  {http://arxiv.org/abs/1105.3865} {arXiv:1105.3865 [nucl-ex]} \BibitemShut
  {NoStop}%
%%CITATION = ARXIV:1105.3865;%%
\bibitem [{\citenamefont {Bhalerao}\ \emph
  {et~al.}(2011{\natexlab{a}})\citenamefont {Bhalerao}, \citenamefont {Luzum},\
  and\ \citenamefont {Ollitrault}}]{Bhalerao:2011ry}%
  \BibitemOpen
  \bibfield  {author} {\bibinfo {author} {\bibfnamefont {R.}~\bibnamefont
  {Bhalerao}}, \bibinfo {author} {\bibfnamefont {M.}~\bibnamefont {Luzum}}, \
  and\ \bibinfo {author} {\bibfnamefont {J.}~\bibnamefont {Ollitrault}},\
  }\href {\doibase 10.1088/0954-3899/38/12/124055} {\bibfield  {journal}
  {\bibinfo  {journal} {J.~Phys.}\ }\textbf {\bibinfo {volume} {G~38}},\ \bibinfo
  {pages} {124055} (\bibinfo {year} {2011}{\natexlab{a}})},\ \Eprint
  {http://arxiv.org/abs/1106.4940} {arXiv:1106.4940 [nucl-ex]} \BibitemShut
  {NoStop}%
%%CITATION = ARXIV:1106.4940;%%
\bibitem [{\citenamefont {Bhalerao}\ \emph
  {et~al.}(2011{\natexlab{b}})\citenamefont {Bhalerao}, \citenamefont {Luzum},\
  and\ \citenamefont {Ollitrault}}]{Bhalerao:2011yg}%
  \BibitemOpen
  \bibfield  {author} {\bibinfo {author} {\bibfnamefont {R.~S.}\ \bibnamefont
  {Bhalerao}}, \bibinfo {author} {\bibfnamefont {M.}~\bibnamefont {Luzum}}, \
  and\ \bibinfo {author} {\bibfnamefont {J.-Y.}\ \bibnamefont {Ollitrault}},\
  }\href {\doibase 10.1103/PhysRevC.84.034910} {\bibfield  {journal} {\bibinfo
  {journal} {Phys.~Rev.}\ }\textbf {\bibinfo {volume} {C~84}},\ \bibinfo {pages}
  {034910} (\bibinfo {year} {2011}{\natexlab{b}})},\ \Eprint
  {http://arxiv.org/abs/1104.4740} {arXiv:1104.4740 [nucl-th]} \BibitemShut
  {NoStop}%
%%CITATION = ARXIV:1104.4740;%%
\bibitem [{\citenamefont {Jia}\ and\ \citenamefont
  {Mohapatra}(2013)}]{Jia:2012ma}%
  \BibitemOpen
  \bibfield  {author} {\bibinfo {author} {\bibfnamefont {J.}~\bibnamefont
  {Jia}}\ and\ \bibinfo {author} {\bibfnamefont {S.}~\bibnamefont
  {Mohapatra}},\ }\href {\doibase 10.1140/epjc/s10052-013-2510-y} {\bibfield
  {journal} {\bibinfo  {journal} {Eur.~Phys.~J.}\ }\textbf {\bibinfo {volume}
  {C~73}},\ \bibinfo {pages} {2510} (\bibinfo {year} {2013})},\ \Eprint
  {http://arxiv.org/abs/1203.5095} {arXiv:1203.5095 [nucl-th]} \BibitemShut
  {NoStop}%
%%CITATION = ARXIV:1203.5095;%%
\bibitem [{\citenamefont {Jia}\ and\ \citenamefont
  {Teaney}(2013)}]{Jia:2012ju}%
  \BibitemOpen
  \bibfield  {author} {\bibinfo {author} {\bibfnamefont {J.}~\bibnamefont
  {Jia}}\ and\ \bibinfo {author} {\bibfnamefont {D.}~\bibnamefont {Teaney}},\
  }\href {\doibase 10.1140/epjc/s10052-013-2558-8} {\bibfield  {journal}
  {\bibinfo  {journal} {Eur.~Phys.~J.}\ }\textbf {\bibinfo {volume} {C~73}},\
  \bibinfo {pages} {2558} (\bibinfo {year} {2013})},\ \Eprint
  {http://arxiv.org/abs/1205.3585} {arXiv:1205.3585 [nucl-ex]} \BibitemShut
  {NoStop}%
%%CITATION = ARXIV:1205.3585;%%
\bibitem [{\citenamefont {Aad}\ \emph {et~al.}(2008)\citenamefont {Aad} \emph
  {et~al.}}]{Aad:2008zzm}%
  \BibitemOpen
  \bibfield  {author} {\bibinfo {author} {\bibfnamefont {ATLAS Collaboration,~}}
  }\href {\doibase 10.1088/1748-0221/3/08/S08003} {\bibfield  {journal}
  {\bibinfo  {journal} {JINST}\ }\textbf {\bibinfo {volume} {3}},\ \bibinfo
  {pages} {S08003} (\bibinfo {year} {2008})}\BibitemShut {NoStop}%
%%CITATION = JINST,3,S08003;%%
\bibitem [{\citenamefont {Aad}\ \emph {et~al.}(2012{\natexlab{b}})\citenamefont
  {Aad} \emph {et~al.}}]{Aad:2012xs}%
  \BibitemOpen
  \bibfield  {author} {\bibinfo {author} {\bibfnamefont {ATLAS Collaboration,~}}
  }\href {\doibase 10.1140/epjc/s10052-011-1849-1} {\bibfield  {journal}
  {\bibinfo  {journal} {Eur.~Phys.~J.}\ }\textbf {\bibinfo {volume} {C~72}},\
  \bibinfo {pages} {1849} (\bibinfo {year} {2012}{\natexlab{b}})},\ \Eprint
  {http://arxiv.org/abs/1110.1530} {arXiv:1110.1530 [hep-ex]} \BibitemShut
  {NoStop}%
%%CITATION = ARXIV:1110.1530;%%
\bibitem [{\citenamefont {Aad}\ \emph {et~al.}(2013{\natexlab{b}})\citenamefont
  {Aad} \emph {et~al.}}]{Aad:2011he}%
  \BibitemOpen
  \bibfield  {author} {\bibinfo {author} {\bibfnamefont {ATLAS Collaboration,~}}
  }\href {\doibase 10.1140/epjc/s10052-013-2304-2} {\bibfield  {journal}
  {\bibinfo  {journal} {Eur.~Phys.~J.}\ }\textbf {\bibinfo {volume} {C~73}},\
  \bibinfo {pages} {2304} (\bibinfo {year} {2013}{\natexlab{b}})},\ \Eprint
  {http://arxiv.org/abs/1112.6426} {arXiv:1112.6426 [hep-ex]} \BibitemShut
  {NoStop}%
%%CITATION = ARXIV:1112.6426;%%
\bibitem [{\citenamefont {Aad}\ \emph {et~al.}(2012{\natexlab{c}})\citenamefont
  {Aad} \emph {et~al.}}]{ATLAS:2011ag}%
  \BibitemOpen
  \bibfield  {author} {\bibinfo {author} {\bibfnamefont {ATLAS Collaboration,~}}
  }\href {\doibase 10.1016/j.physletb.2012.02.045} {\bibfield  {journal}
  {\bibinfo  {journal} {Phys.~Lett.}\ }\textbf {\bibinfo {volume} {B~710}},\
  \bibinfo {pages} {363} (\bibinfo {year} {2012}{\natexlab{c}})},\ \Eprint
  {http://arxiv.org/abs/1108.6027} {arXiv:1108.6027 [hep-ex]} \BibitemShut
  {NoStop}%
%%CITATION = ARXIV:1108.6027;%%
\bibitem [{\citenamefont {Aad}\ \emph {et~al.}(2011)\citenamefont {Aad} \emph
  {et~al.}}]{Aad:2010ac}%
  \BibitemOpen
  \bibfield  {author} {\bibinfo {author} {\bibfnamefont {ATLAS Collaboration,~}}
  }\href {\doibase 10.1088/1367-2630/13/5/053033} {\bibfield  {journal}
  {\bibinfo  {journal} {New~J.~Phys.}\ }\textbf {\bibinfo {volume} {13}},\
  \bibinfo {pages} {053033} (\bibinfo {year} {2011})},\ \Eprint
  {http://arxiv.org/abs/1012.5104} {arXiv:1012.5104 [hep-ex]} \BibitemShut
  {NoStop}%
%%CITATION = ARXIV:1012.5104;%%
\bibitem [{\citenamefont {Aad}\ \emph {et~al.}(2012{\natexlab{d}})\citenamefont
  {Aad} \emph {et~al.}}]{ATLAS:2011ah}%
  \BibitemOpen
  \bibfield  {author} {\bibinfo {author} {\bibfnamefont  {ATLAS Collaboration,~}}}\href {\doibase
  10.1016/j.physletb.2011.12.056} {\bibfield  {journal} {\bibinfo  {journal}
  {Phys.~Lett.}\ }\textbf {\bibinfo {volume} {B~707}},\ \bibinfo {pages} {330}
  (\bibinfo {year} {2012}{\natexlab{d}})},\ \Eprint
  {http://arxiv.org/abs/1108.6018} {arXiv:1108.6018 [hep-ex]} \BibitemShut
  {NoStop}%
%%CITATION = ARXIV:1108.6018;%%
\bibitem [{\citenamefont {Gyulassy}\ and\ \citenamefont
  {Wang}(1994)}]{Gyulassy:1994ew}%
  \BibitemOpen
  \bibfield  {author} {\bibinfo {author} {\bibfnamefont {M.}~\bibnamefont
  {Gyulassy}}\ and\ \bibinfo {author} {\bibfnamefont {X.-N.}\ \bibnamefont
  {Wang}},\ }\href {\doibase 10.1016/0010-4655(94)90057-4} {\bibfield
  {journal} {\bibinfo  {journal} {Comput.~Phys.~Commun.}\ }\textbf {\bibinfo
  {volume} {83}},\ \bibinfo {pages} {307} (\bibinfo {year} {1994})},\ \Eprint
  {http://arxiv.org/abs/nucl-th/9502021} {arXiv:nucl-th/9502021 [nucl-th]}
  \BibitemShut {NoStop}%
%%CITATION = NUCL-TH/9502021;%%
\bibitem [{\citenamefont {{ATLAS Collaboration}}(2010)}]{Aad:2010ah}%
  \BibitemOpen
  \bibfield  {author} {\bibinfo {author} {\bibnamefont {{ATLAS
  Collaboration}}},\ }\href {\doibase 10.1140/epjc/s10052-010-1429-9}
  {\bibfield  {journal} {\bibinfo  {journal} {Eur.~Phys.~J.}\ }\textbf
  {\bibinfo {volume} {C~70}},\ \bibinfo {pages} {823} (\bibinfo {year}
  {2010})}\BibitemShut {NoStop}%
\bibitem [{\citenamefont {Agostinelli}\ \emph {et~al.}(2003)\citenamefont
  {Agostinelli} \emph {et~al.}}]{Agostinelli:2002hh}%
  \BibitemOpen
  \bibfield  {author} {\bibinfo {author} {\bibfnamefont {S.}~\bibnamefont
  {Agostinelli}} \emph {et~al.} (\bibinfo {collaboration} {GEANT4}),\ }\href
  {\doibase 10.1016/S0168-9002(03)01368-8} {\bibfield  {journal} {\bibinfo
  {journal} {Nucl.~Instrum.~Methods~Phys.~Res.,~Sect.~}\ }\textbf {\bibinfo {volume} {A~506}},\
  \bibinfo {pages} {250} (\bibinfo {year} {2003})}\BibitemShut {NoStop}%
%%CITATION = NUIMA,A506,250;%%
\bibitem [{trk()}]{trk}%
  \BibitemOpen
  \href@noop {} {}\bibinfo {note} {{ ATLAS} Collaboration,
  ATLAS-CONF-2011-079, http://cdsweb.cern.ch/record/1355702}\BibitemShut
  {NoStop}%
\bibitem [{\citenamefont {Afanasiev}\ \emph {et~al.}(2009)\citenamefont
  {Afanasiev} \emph {et~al.}}]{Afanasiev:2009wq}%
  \BibitemOpen
  \bibfield  {author} {\bibinfo {author} {\bibfnamefont {S.}~\bibnamefont
  {Afanasiev}} \emph {et~al.} (\bibinfo {collaboration} {PHENIX
  Collaboration}),\ }\href {\doibase 10.1103/PhysRevC.80.024909} {\bibfield
  {journal} {\bibinfo  {journal} {Phys.~Rev.}\ }\textbf {\bibinfo {volume}
  {C~80}},\ \bibinfo {pages} {024909} (\bibinfo {year} {2009})},\ \Eprint
  {http://arxiv.org/abs/0905.1070} {arXiv:0905.1070 [nucl-ex]} \BibitemShut
  {NoStop}%
%%CITATION = ARXIV:0905.1070;%%
\bibitem [{\citenamefont {Barrette}\ \emph {et~al.}(1997)\citenamefont
  {Barrette} \emph {et~al.}}]{Barrette:1997pt}%
  \BibitemOpen
  \bibfield  {author} {\bibinfo {author} {\bibfnamefont {J.}~\bibnamefont
  {Barrette}} \emph {et~al.} (\bibinfo {collaboration} {E877 Collaboration}),\
  }\href {\doibase 10.1103/PhysRevC.56.3254} {\bibfield  {journal} {\bibinfo
  {journal} {Phys.~Rev.}\ }\textbf {\bibinfo {volume} {C~56}},\ \bibinfo {pages}
  {3254} (\bibinfo {year} {1997})},\ \Eprint
  {http://arxiv.org/abs/nucl-ex/9707002} {arXiv:nucl-ex/9707002 [nucl-ex]}
  \BibitemShut {NoStop}%
%%CITATION = NUCL-EX/9707002;%%
\bibitem [{\citenamefont {Luzum}\ and\ \citenamefont
  {Ollitrault}(2013)}]{Luzum:2012da}%
  \BibitemOpen
  \bibfield  {author} {\bibinfo {author} {\bibfnamefont {M.}~\bibnamefont
  {Luzum}}\ and\ \bibinfo {author} {\bibfnamefont {J.-Y.}\ \bibnamefont
  {Ollitrault}},\ }\href {\doibase 10.1103/PhysRevC.87.044907} {\bibfield
  {journal} {\bibinfo  {journal} {Phys.~Rev.}\ }\textbf {\bibinfo {volume}
  {C~87}},\ \bibinfo {pages} {044907} (\bibinfo {year} {2013})},\ \Eprint
  {http://arxiv.org/abs/1209.2323} {arXiv:1209.2323 [nucl-ex]} \BibitemShut
  {NoStop}%
%%CITATION = ARXIV:1209.2323;%%
\bibitem [{\citenamefont {Bhalerao}\ \emph {et~al.}(2013)\citenamefont
  {Bhalerao}, \citenamefont {Ollitrault},\ and\ \citenamefont
  {Pal}}]{Bhalerao:2013ina}%
  \BibitemOpen
  \bibfield  {author} {\bibinfo {author} {\bibfnamefont {R.~S.}\ \bibnamefont
  {Bhalerao}}, \bibinfo {author} {\bibfnamefont {J.-Y.}\ \bibnamefont
  {Ollitrault}}, \ and\ \bibinfo {author} {\bibfnamefont {S.}~\bibnamefont
  {Pal}},\ }\href {\doibase 10.1103/PhysRevC.88.024909} {\bibfield  {journal}
  {\bibinfo  {journal} {Phys.~Rev.}\ }\textbf {\bibinfo {volume}
  {C~88}},\ \bibinfo {pages} {024909} (\bibinfo {year} {2013})},\ \Eprint
  {http://arxiv.org/abs/1307.0980} {arXiv:1307.0980 [nucl-th]} \BibitemShut
  {NoStop}%
%%CITATION = ARXIV:1307.0980;%%
\bibitem [{\citenamefont {Adler}\ \emph {et~al.}(2002)\citenamefont {Adler}
  \emph {et~al.}}]{Adler:2002pu}%
  \BibitemOpen
  \bibfield  {author} {\bibinfo {author} {\bibfnamefont {C.}~\bibnamefont
  {Adler}} \emph {et~al.} (\bibinfo {collaboration} {STAR Collaboration}),\
  }\href {\doibase 10.1103/PhysRevC.66.034904} {\bibfield  {journal} {\bibinfo
  {journal} {Phys.~Rev.}\ }\textbf {\bibinfo {volume} {C~66}},\ \bibinfo {pages}
  {034904} (\bibinfo {year} {2002})},\ \Eprint
  {http://arxiv.org/abs/nucl-ex/0206001} {arXiv:nucl-ex/0206001 [nucl-ex]}
  \BibitemShut {NoStop}%
%%CITATION = NUCL-EX/0206001;%%
\bibitem [{\citenamefont {Andrade}\ \emph {et~al.}(2011)\citenamefont
  {Andrade}, \citenamefont {Grassi}, \citenamefont {Hama},\ and\ \citenamefont
  {Qian}}]{Andrade:2010sd}%
  \BibitemOpen
  \bibfield  {author} {\bibinfo {author} {\bibfnamefont {R.}~\bibnamefont
  {Andrade}}, \bibinfo {author} {\bibfnamefont {F.}~\bibnamefont {Grassi}},
  \bibinfo {author} {\bibfnamefont {Y.}~\bibnamefont {Hama}}, \ and\ \bibinfo
  {author} {\bibfnamefont {W.-L.}\ \bibnamefont {Qian}},\ }\href {\doibase
  10.1016/j.nuclphysa.2010.08.004} {\bibfield  {journal} {\bibinfo  {journal}
  {Nucl.~Phys.~A}\ }\textbf {\bibinfo {volume} {854}},\ \bibinfo {pages} {81}
  (\bibinfo {year} {2011})},\ \Eprint {http://arxiv.org/abs/1008.0139}
  {arXiv:1008.0139 [hep-ph]} \BibitemShut {NoStop}%
%%CITATION = ARXIV:1008.0139;%%
\bibitem [{\citenamefont {Niemi}\ \emph {et~al.}(2013)\citenamefont {Niemi},
  \citenamefont {Denicol}, \citenamefont {Holopainen},\ and\ \citenamefont
  {Huovinen}}]{Niemi:2012aj}%
  \BibitemOpen
  \bibfield  {author} {\bibinfo {author} {\bibfnamefont {H.}~\bibnamefont
  {Niemi}}, \bibinfo {author} {\bibfnamefont {G.}~\bibnamefont {Denicol}},
  \bibinfo {author} {\bibfnamefont {H.}~\bibnamefont {Holopainen}}, \ and\
  \bibinfo {author} {\bibfnamefont {P.}~\bibnamefont {Huovinen}},\ }\href
  {\doibase 10.1103/PhysRevC.87.054901} {\bibfield  {journal} {\bibinfo
  {journal} {Phys.~Rev.}\ }\textbf {\bibinfo {volume} {C~87}},\ \bibinfo {pages}
  {054901} (\bibinfo {year} {2013})},\ \Eprint {http://arxiv.org/abs/1212.1008}
  {arXiv:1212.1008 [nucl-th]} \BibitemShut {NoStop}%
%%CITATION = ARXIV:1212.1008;%%
\bibitem [{\citenamefont {Collaboration}(2013)}]{epcorr}%
  \BibitemOpen
  \bibfield  {author} {\bibinfo {author} {\bibfnamefont {ATLAS Collaboration,}}\ }
%\href {\doibase 10.1016/j.nuclphysa.2012.12.043}
%  {\bibfield  {journal} {\bibinfo  {journal} {Nucl.~Phys.~A}\ }\textbf
%  {\bibinfo {volume} {910}},\ \bibinfo {pages} {276} (\bibinfo {year}  {2013})},\ 
\bibinfo {note} {{ATLAS-CONF-2012-049},  https://cds.cern.ch/record/1451882}
\BibitemShut {NoStop}%
\bibitem [{\citenamefont {Teaney}\ and\ \citenamefont
  {Yan}(2012)}]{Teaney:2012ke}%
  \BibitemOpen
  \bibfield  {author} {\bibinfo {author} {\bibfnamefont {D.}~\bibnamefont
  {Teaney}}\ and\ \bibinfo {author} {\bibfnamefont {L.}~\bibnamefont {Yan}},\
  }\href {\doibase 10.1103/PhysRevC.86.044908} {\bibfield  {journal} {\bibinfo
  {journal} {Phys.~Rev.}\ }\textbf {\bibinfo {volume} {C~86}},\ \bibinfo {pages}
  {044908} (\bibinfo {year} {2012})},\ \Eprint {http://arxiv.org/abs/1206.1905}
  {arXiv:1206.1905 [nucl-th]} \BibitemShut {NoStop}%
%%CITATION = ARXIV:1206.1905;%%
\bibitem [{\citenamefont {Qiu}\ and\ \citenamefont {Heinz}(2012)}]{Qiu:2012uy}%
  \BibitemOpen
  \bibfield  {author} {\bibinfo {author} {\bibfnamefont {Z.}~\bibnamefont
  {Qiu}}\ and\ \bibinfo {author} {\bibfnamefont {U.}~\bibnamefont {Heinz}},\
  }\href {\doibase 10.1016/j.physletb.2012.09.030} {\bibfield  {journal}
  {\bibinfo  {journal} {Phys.~Lett.}\ }\textbf {\bibinfo {volume} {B~717}},\
  \bibinfo {pages} {261} (\bibinfo {year} {2012})},\ \Eprint
  {http://arxiv.org/abs/1208.1200} {arXiv:1208.1200 [nucl-th]} \BibitemShut
  {NoStop}%
%%CITATION = ARXIV:1208.1200;%%
\bibitem [{\citenamefont {Teaney}\ and\ \citenamefont
  {Yan}(2013{\natexlab{a}})}]{Teaney:2012gu}%
  \BibitemOpen
  \bibfield  {author} {\bibinfo {author} {\bibfnamefont {D.}~\bibnamefont
  {Teaney}}\ and\ \bibinfo {author} {\bibfnamefont {L.}~\bibnamefont {Yan}},\
  }\href {\doibase 10.1016/j.nuclphysa.2013.02.025} {\bibfield  {journal}
  {\bibinfo  {journal} {Nucl.~Phys.}\ }\textbf {\bibinfo {volume}
  {A~904}},\ \bibinfo {pages} {365c} (\bibinfo {year} {2013}{\natexlab{a}})},\
  \Eprint {http://arxiv.org/abs/1210.5026} {arXiv:1210.5026 [nucl-th]}
  \BibitemShut {NoStop}%
%%CITATION = ARXIV:1210.5026;%%
\bibitem [{\citenamefont {Teaney}\ and\ \citenamefont
  {Yan}(2013{\natexlab{b}})}]{Teaney:2013dta}%
  \BibitemOpen
  \bibfield  {author} {\bibinfo {author} {\bibfnamefont {D.}~\bibnamefont
  {Teaney}}\ and\ \bibinfo {author} {\bibfnamefont {L.}~\bibnamefont {Yan}},\
  }\ \Eprint
  {http://arxiv.org/abs/1312.3689} {arXiv:1312.3689 [nucl-th]} \BibitemShut
  {NoStop}%
%%CITATION = ARXIV:1312.3689;%%
\bibitem [{\citenamefont {Lin}\ \emph {et~al.}(2005)\citenamefont {Lin},
  \citenamefont {Ko}, \citenamefont {Li}, \citenamefont {Zhang},\ and\
  \citenamefont {Pal}}]{Lin:2004en}%
  \BibitemOpen
  \bibfield  {author} {\bibinfo {author} {\bibfnamefont {Z.-W.}\ \bibnamefont
  {Lin}}, \bibinfo {author} {\bibfnamefont {C.~M.}\ \bibnamefont {Ko}},
  \bibinfo {author} {\bibfnamefont {B.-A.}\ \bibnamefont {Li}}, \bibinfo
  {author} {\bibfnamefont {B.}~\bibnamefont {Zhang}}, \ and\ \bibinfo {author}
  {\bibfnamefont {S.}~\bibnamefont {Pal}},\ }\href {\doibase
  10.1103/PhysRevC.72.064901} {\bibfield  {journal} {\bibinfo  {journal}
  {Phys.~Rev.}\ }\textbf {\bibinfo {volume} {C~72}},\ \bibinfo {pages} {064901}
  (\bibinfo {year} {2005})},\ \Eprint {http://arxiv.org/abs/nucl-th/0411110}
  {arXiv:nucl-th/0411110 [nucl-th]} \BibitemShut {NoStop}%
%%CITATION = NUCL-TH/0411110;%%
\bibitem [{\citenamefont {Xu}\ and\ \citenamefont
  {Ko}(2011{\natexlab{a}})}]{Xu:2011jm}%
  \BibitemOpen
  \bibfield  {author} {\bibinfo {author} {\bibfnamefont {J.}~\bibnamefont
  {Xu}}\ and\ \bibinfo {author} {\bibfnamefont {C.~M.}\ \bibnamefont {Ko}},\
  }\href {\doibase 10.1103/PhysRevC.84.044907} {\bibfield  {journal} {\bibinfo
  {journal} {Phys.~Rev.}\ }\textbf {\bibinfo {volume} {C~84}},\ \bibinfo {pages}
  {044907} (\bibinfo {year} {2011}{\natexlab{a}})},\ \Eprint
  {http://arxiv.org/abs/1108.0717} {arXiv:1108.0717 [nucl-th]} \BibitemShut
  {NoStop}%
%%CITATION = ARXIV:1108.0717;%%
\bibitem [{\citenamefont {Xu}\ and\ \citenamefont
  {Ko}(2011{\natexlab{b}})}]{Xu:2011fi}%
  \BibitemOpen
  \bibfield  {author} {\bibinfo {author} {\bibfnamefont {J.}~\bibnamefont
  {Xu}}\ and\ \bibinfo {author} {\bibfnamefont {C.~M.}\ \bibnamefont {Ko}},\
  }\href {\doibase 10.1103/PhysRevC.83.034904} {\bibfield  {journal} {\bibinfo
  {journal} {Phys.~Rev.}\ }\textbf {\bibinfo {volume} {C~83}},\ \bibinfo {pages}
  {034904} (\bibinfo {year} {2011}{\natexlab{b}})},\ \Eprint
  {http://arxiv.org/abs/1101.2231} {arXiv:1101.2231 [nucl-th]} \BibitemShut
  {NoStop}%
%%CITATION = ARXIV:1101.2231;%%
\end{thebibliography}
%\bibliographystyle{apsrev4-1}

%\begin{thebibliography}{29} % do not change

\section*{Appendix}
\begin{figure}[h!]
\begin{center}
\includegraphics[width=1\columnwidth]{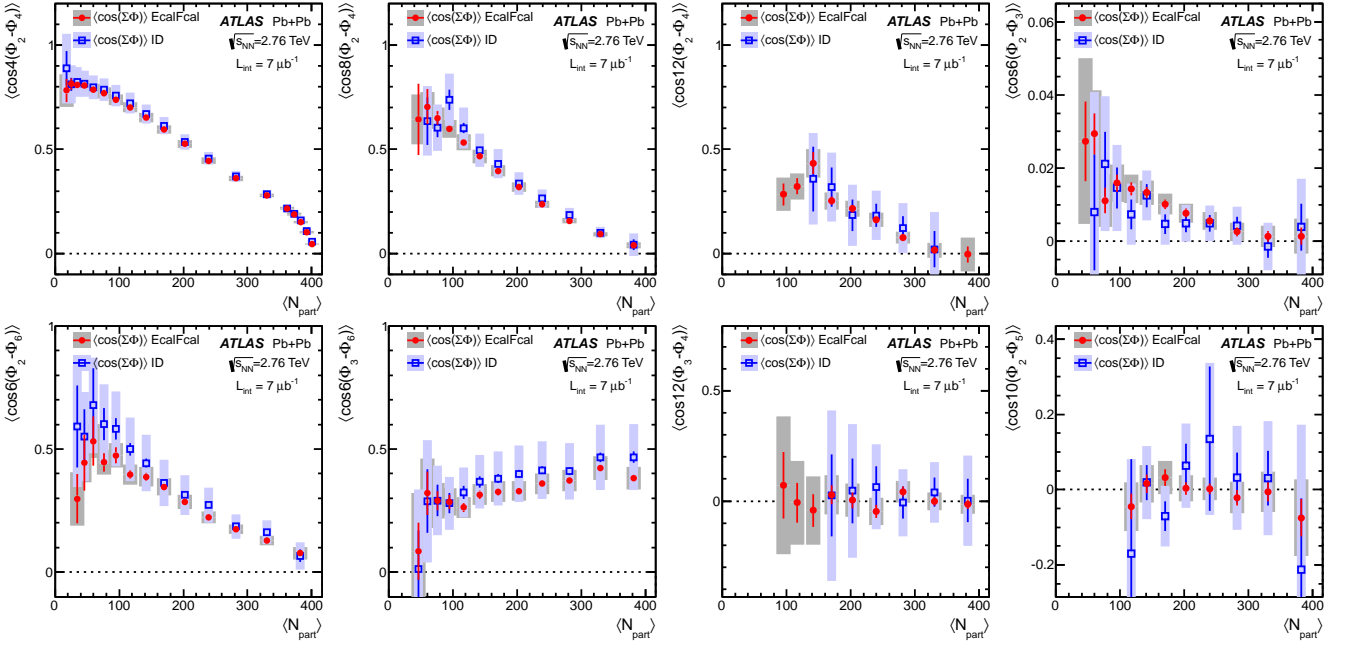}
\end{center}
\caption{\label{fig:comp2p} (Color online) The comparison of the eight two-plane correlators between the calorimeters (default) and ID (cross-check) as a function of $\langle N_{\mathrm{part}}\rangle$, both obtained from the EP method. The error bars and the shaded bands indicate the statistical and total systematic uncertainties, respectively.}
\end{figure}

\begin{figure}[h!]
\begin{center}
\includegraphics[width=1\columnwidth]{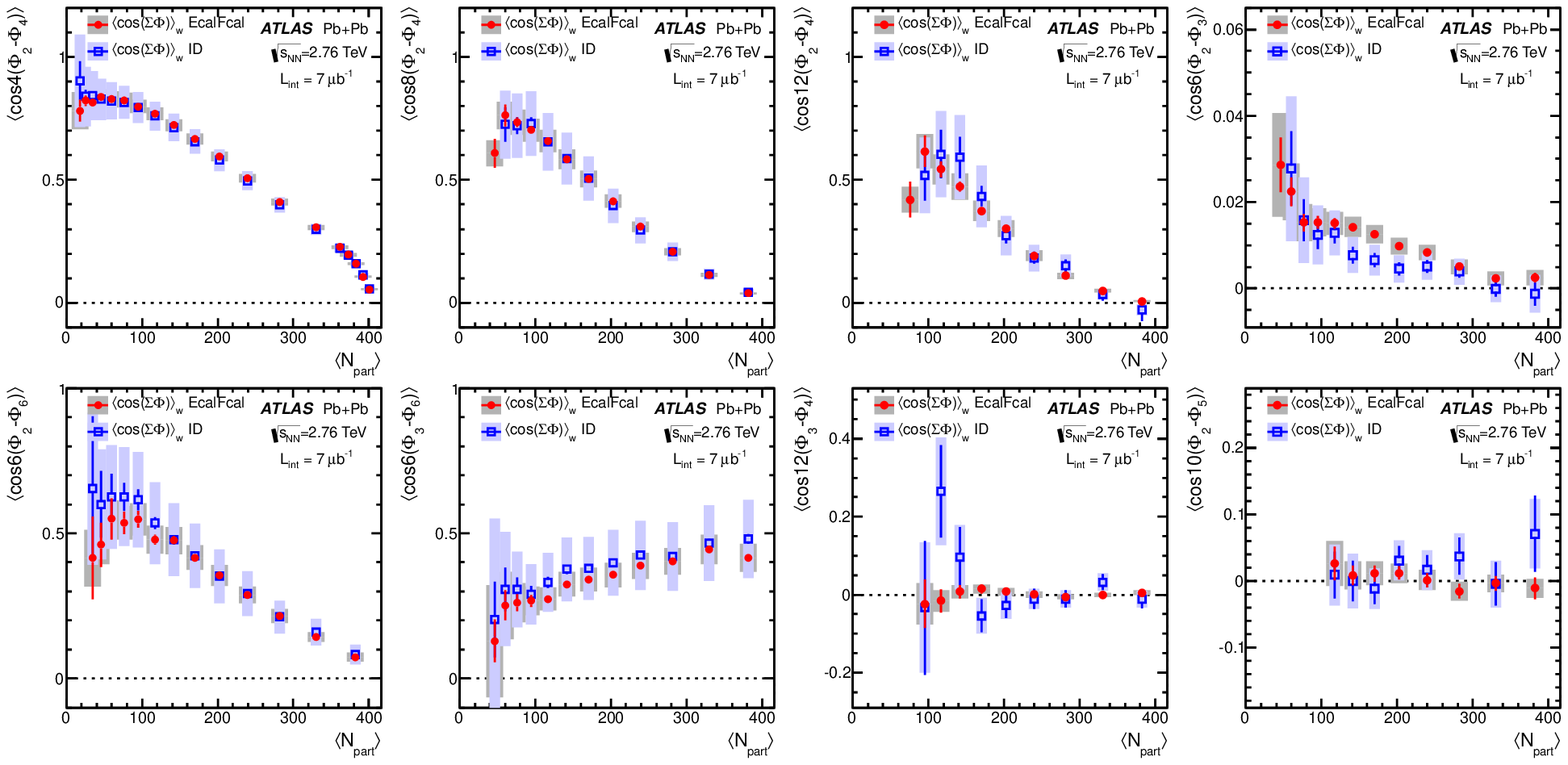}
\end{center}
\caption{\label{fig:comp2pb} The comparison of the eight two-plane correlators between the calorimeters (default) and ID (cross-check) as a function of $\langle N_{\mathrm{part}}\rangle$, both obtained from the SP method.  The error bars and the shaded bands indicate the statistical and total systematic uncertainties, respectively.}
\end{figure}

\begin{figure}[h!]
\begin{center}
\includegraphics[width=0.8\columnwidth]{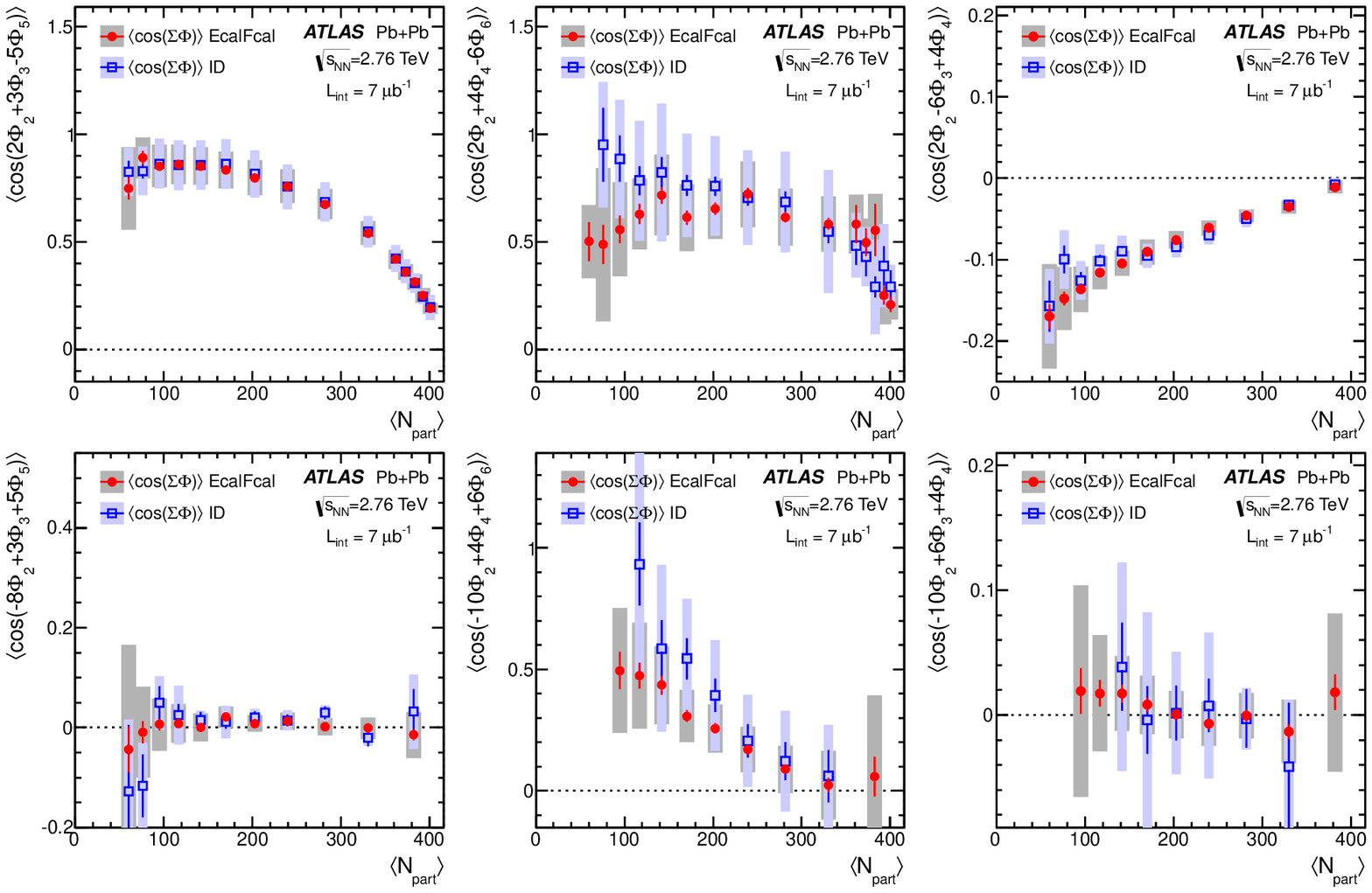}
\end{center}
\caption{\label{fig:comp3p} (Color online) The comparison of the six three-plane correlators between the calorimeters (default) and ID (cross-check) as a function of $\langle N_{\mathrm{part}}\rangle$, both obtained from the EP method. The error bars and the shaded bands indicate the statistical and total systematic uncertainty, respectively.}
\end{figure}

\begin{figure}[h!]
\begin{center}
\includegraphics[width=0.8\columnwidth]{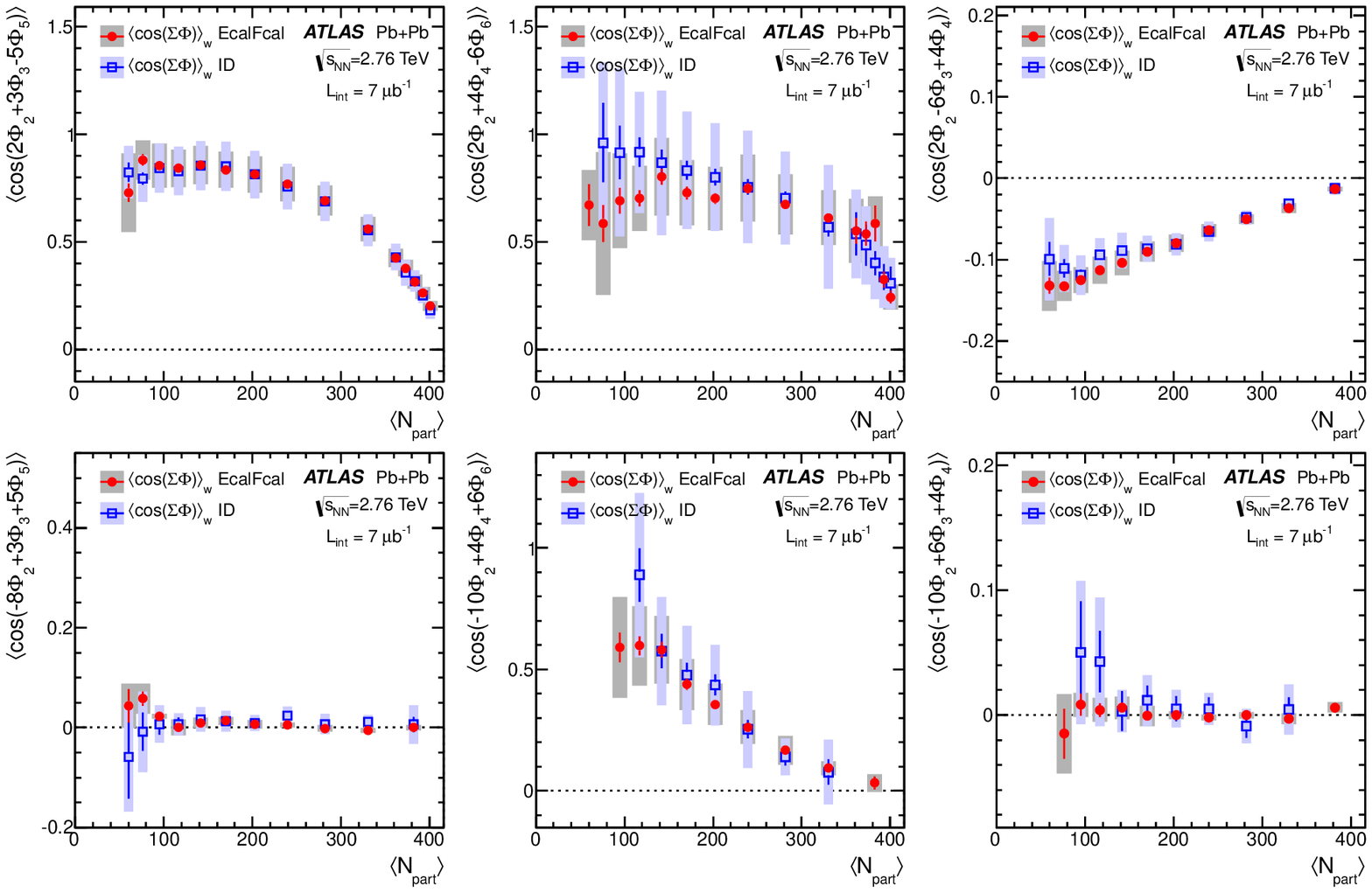}
\end{center}
\caption{\label{fig:comp3pb} (Color online) The comparison of the six three-plane correlators between the calorimeters (default) and ID (cross-check) as a function of $\langle N_{\mathrm{part}}\rangle$, both obtained from the SP method.  The error bars and the shaded bands indicate the statistical and total systematic uncertainties, respectively.}
\end{figure}

Figures~\ref{fig:comp2p}--\ref{fig:comp3pb} compare results between the calorimeter and the ID for the two-plane and three-plane correlations. As discussed at the end of Sec.~\ref{sec:sys}, the results are consistent between the calorimeter and the ID within their respective systematic uncertainties. 

\clearpage 
% ATLAS Collaboration author list
% Data extracted on 22-May-2014 for paper reference HION-2012-03
%\documentclass[11pt]{article}
%\usepackage{a4wide}\begin{document}
\begin{flushleft}
{\Large The ATLAS Collaboration}

\bigskip

G.~Aad$^{\rm 84}$,
B.~Abbott$^{\rm 112}$,
J.~Abdallah$^{\rm 152}$,
S.~Abdel~Khalek$^{\rm 116}$,
O.~Abdinov$^{\rm 11}$,
R.~Aben$^{\rm 106}$,
B.~Abi$^{\rm 113}$,
M.~Abolins$^{\rm 89}$,
O.S.~AbouZeid$^{\rm 159}$,
H.~Abramowicz$^{\rm 154}$
% ...
\bigskip
\\
% ...
$^{11}$ Institute of Physics, Azerbaijan Academy of Sciences, Baku, Azerbaijan\\
% ...
$^{84}$ CPPM, Aix-Marseille Universit{\'e} and CNRS/IN2P3, Marseille, France\\
% ...
$^{89}$ Department of Physics and Astronomy, Michigan State University, East Lansing MI, United States of America\\
% ...
$^{106}$ Nikhef National Institute for Subatomic Physics and University of Amsterdam, Amsterdam, Netherlands\\
% ...
$^{112}$ Homer L. Dodge Department of Physics and Astronomy, University of Oklahoma, Norman OK, United States of America\\
$^{113}$ Department of Physics, Oklahoma State University, Stillwater OK, United States of America\\
% ...
$^{116}$ LAL, Universit{\'e} Paris-Sud and CNRS/IN2P3, Orsay, France\\
% ...
$^{152}$ Institute of Physics, Academia Sinica, Taipei, Taiwan\\
% ...
$^{154}$ Raymond and Beverly Sackler School of Physics and Astronomy, Tel Aviv University, Tel Aviv, Israel\\
% ...
$^{159}$ Department of Physics, University of Toronto, Toronto ON, Canada\\
% ...
\end{flushleft}

%\end{document}
% Created with ./xml2latex.py
 
\end{document}